\documentclass[%
aps,
preprint,
nofootinbib,
eqsecnum,
a4paper,
showpacs,
 prd,
]{revtex4-2}
\usepackage{amsmath}
\usepackage{amssymb}
\usepackage{slashed}
\usepackage{mathrsfs}
\usepackage{braket}
\usepackage{graphicx}

\newcommand{\threeflavor}{{\begin{pmatrix}{\nu_e}\\ {\nu_{\mu}}\\ {\nu_{\tau}}\end{pmatrix}}}

\begin{document}
	
\title{Effect of spacetime geometry on neutrino oscillations}
\author{Riya Barick}
\email{riyabarik7@gmail.com}
\author{Indrajit Ghose}
\email{ghose.meghnad@gmail.com}
\author{Amitabha Lahiri}
\email{amitabha@bose.res.in}
\affiliation{S. N. Bose National Centre for Basic Sciences\\
	Block JD, Sector 3, Salt Lake, WB 700106, INDIA.}

\begin{abstract}
Propagation of fermions in spacetime requires a spin connection, which can be split into a universal gravitational part and a non-universal ``contorsion'' part. 
The latter is non-dynamical and can be eliminated from the theory, leaving an effective four-fermion interaction with unknown coupling constants.  
The generic form of the contorsion-fermion coupling, and thus the four-fermion interaction, breaks chiral symmetry. This interaction affects all 
fermions --- in particular neutrinos passing through matter will notice a contribution to their effective Hamiltonian, just like the MSW effect 
coming from weak interactions, but diagonal in the mass basis rather than the flavor basis. Then there is a possibility that this geometrical 
contribution is not negligible for neutrinos passing through normal matter, provided the coupling constants are not too small. We calculate the 
matter potential due to this interaction and thus write the conversion and survival probabilities including its effect. We plot conversion probabilities of $\nu_\mu$ to $\nu_e$ and $\nu_\tau$ for a baseline of 1300 km, as well as their dependence on the CP phase, with and without the geometrical interaction. We also plot the survival probability of reactor $\bar{\nu}_e$ for different baselines.

\end{abstract}

\maketitle
\newpage

\section{Introduction}
Neutrinos in the Standard Model are massless and purely left-handed, described by two-component Weyl spinors. There is no right-handed neutrino in the Standard Model, as right-handed fermions are singlets of SU(2)$_L$ and since neutrinos are colorless and electrically neutral, right-handed neutrinos are singlets of the full Standard Model gauge group and are thus must be decoupled from the rest of the theory. Absence of right-handed neutrinos means that there is no Dirac mass term, and neutrinos do not have a Majorana mass term because there is no Higgs triplet. On the other hand, the observation of disappearance (and in some cases, appearance) of flavors among atmospheric neutrinos~\cite{Super-Kamiokande:1998kpq}, solar neutrinos~\cite{SNO:2002tuh}, reactor neutrinos~\cite{KamLAND:2002uet}, accelerator neutrinos~\cite{T2K:2011ypd}, and in many other experiments~\cite{K2K:2002icj,MINOS:2006foh,DayaBay:2012fng, DoubleChooz:2011ymz, RENO:2012mkc} point to neutrino flavor mixing and oscillation, which is explained by the existence of neutrino masses. This clearly indicates the existence of physics beyond the Standard Model. 

There are many ways of extending the Standard Model to include a massive neutrino~\cite{Mohapatra:1998rq}. The minimal extension of the Standard Model to do this is to keep the gauge group unchanged and either add a right-handed component of the neutrino so that there is a mass term after the electroweak symmetry breaking, or to allow nonconservation of lepton number (more precisely, $B-L$) and have only left-handed neutrinos, albeit with a Majorana mass term. Different types of mass terms will have different consequences even for neutrino oscillations -- precision measurements now allow determination of various parameters like mixing angles and mass squared differences to an accuracy of a few per cent. Neutrino oscillations thus provide a glimpse of the physics beyond the Standard Model~\cite{Chatterjee:2015gta,Gouttenoire:2022gwi, Barranco:2007tz,Antusch:2008tz}.

It goes without saying that before venturing into the unknown domain beyond the Standard Model, we should ensure that all known physics has been taken into account. Although the effects of Standard Model gauge fields on neutrino interactions are well studied, there is one sector of low energy physics which is rarely discussed in the context of neutrino oscillations, and that is gravitation -- more precisely, spacetime geometry. Since gravity is a very weak force, it is generally thought that it will not contribute to neutrino processes. { However, it has long been suspected that the interplay of quantum mechanics with  spacetime geometry may have an effect on neutrino oscillations. The gravitationally induced quantum mechanical phases discussed in~\cite{Ahluwalia:1996ev,Kojima:1996vb,Fornengo:1996ef,Capozziello:1999qm,Cardall:1996cd} or the Berry phase induced by torsion, discussed in~\cite{Capozziello:2000ue} are examples of this, as is the neutrino oscillation near black holes induced by a scalar conformally coupled to gravity~\cite{Mastrototaro:2021kmw}. Another class of examples come from considering the coupling of fermions to spacetime torsion~\cite{Adak:2000tp, Fabbri:2015jaa, DeSabbata:1981ek, Zhang:2000ue, Capozziello:1999nt}.
	Yet another possibility was suggested recently that dynamically generated spacetime torsion could provide another example where geometry has an observable effect on neutrino oscillations~\cite{Chakrabarty:2019cau}.
}
The purpose of this paper is to make that suggestion more concrete, and possibly testable, by calculating such effects on neutrino conversion and survival probabilities. 

The layout of the paper is as follows. We start with a very brief review of fermions in curved spacetime in Sec.~\ref{gravity} and then show how a 4-fermion interaction arises from it. In Sec.~\ref{2nu} we calculate the modified mixing matrix for two neutrino flavors, and then we do the same for three neutrino flavors in Sec.~\ref{3nu}. We end with a discussion of the results in Sec.~\ref{disc}.

\section{Fermions in spacetime}\label{gravity}
It is a basic tenet of General Relativity that the presence of matter changes the geometry of spacetime. Thus the equations of motion of any field must include terms coming from the spacetime geometry. For scalars and gauge bosons, the Standard Model bosonic fields, these are curvature terms which appear in the d'Alembertian and can be neglected in general when the curvature of spacetime can be taken to be small, as in most phenomena in the universe. The role of spacetime in the dynamics of fermions is more complicated. As we will show in this section, spacetime geometry causes an effective four-fermion interaction, with coupling constants which are not universal and may be quite arbitrary, fixed fully by experimental observations.

Fermions in curved spacetime are most conveniently described in the Einstein-Cartan-Sciama-Kibble (ECSK) formalism, a first order formulation of gravity which uses tetrad fields~\cite{Kibble:1961ba,Sciama:1964wt,Hehl:1976kj, Cartan:1923zea, Cartan:1924yea}. Here we give a very brief overview to set up our notations, interested readers can find more details about the ECSK formulation in the literature~\cite{Gasperini:2013, Chakrabarty:2018ybk, Hehl:1974cn, Hehl:2007bn, Hammond:2002rm, Mielke:2017nwt, Poplawski:2009fb}.

In this approach, gravity is written essentially as a gauge theory of the Poincar\' e group. Dirac's $\gamma$ matrices are defined on an ``internal'' flat space, isomorphic to the tangent space at each point, by $\left[\gamma^a, \gamma^b\right]_{+} = 2\eta^{ab}$\,. Tetrad fields  $e^\mu_a$ and their inverses, the co-tetrads $e^a_\mu$\,, relate the spacetime with the internal flat space through 
\begin{align}
	\eta_{{ab}}e^a_\mu e^b_\nu = g_{\mu\nu}\,, \quad g_{\mu\nu}e^\mu_a e^\nu_b = \eta_{ab}\,,
	 \quad e^\mu_a e^a_\nu = \delta^\mu_\nu\,.
\end{align}
We will denote the spacetime indices by lowercase Greek letters $\mu, \nu, \lambda, \cdots$ and internal flat space indices by lowercase Latin letters $a, b, c, \cdots$\,. The spacetime $\gamma$ matrices $\gamma^\mu := e^\mu_a \gamma^a$ satisfy $\left[\gamma_\mu\,, \gamma_\nu\right]_+ = 2g_{\mu\nu}$\,; we will sometimes use $\gamma^\mu$ if the combination $e^\mu_a \gamma^a$ appears in an equation. The co-tetrad, thought of as a 4$\times$4 matrix, has determinant equal to the square root of the matrix determinant,
$|e| = \sqrt{|g|}$\,.  The connection has two parts, one for the spacetime and one for the internal space, and it must be tetrad compatible if it is to annihilate the spacetime $\gamma$ matrices, meaning that 
\begin{align}
&	\nabla_\mu \gamma_\nu = 0 \qquad \Rightarrow \quad \nabla_\mu e_\nu^a = 0 \notag \\
\Rightarrow \qquad &	e^\lambda_a \partial_\mu e^a_\nu + 	A_{\mu}{}^{a}{}_{b} e^b_\nu e^\lambda_a - \Gamma^\lambda{}_{\mu \nu} = 0\,.
	\label{tetrad-postulate}
\end{align}
This relation is often assumed a priori and given the name \textit{tetrad postulate}. The object $A_{\mu}{}^{a}{}_{b}$ is the \textit{spin connection}, which appears in the covariant derivative of spinor fields in minimal substitution,
\begin{equation}\label{Dirac-operator}
	D_\mu\psi = \partial_\mu\psi -\frac{i}{4} A_\mu{}^{ab} \sigma_{ab}\psi\,, \qquad \sigma_{ab} = \frac{i}{2}\left[\gamma_a\,, \gamma_b\right]_{-}\,.
\end{equation}
Note that in this section we are using the signature $(-+++)$\,.

The connection components $\Gamma^{\lambda}{}_{\mu \nu}$ are not a priori assumed to be symmetric -- the connection is not torsion-free in presence of fermions. Let us therefore split the spin connection as 
\begin{equation}\label{split}
	A_{\mu}{}^{ab} = \omega_{\mu}{}^{ab}  + \Lambda_{\mu}{}^{ab}\,,
\end{equation}
where $\omega_{\mu}{}^{ab}$ corresponds to the torsion-free Levi-Civita connection and $\Lambda_{\mu}{}^{ab}\,$ is called \textit{contorsion}.
If $\omega_{\mu}{}^{ab}$ is inserted into Eq.~(\ref{tetrad-postulate}) in place of $	A_{\mu}{}^{ab}$\,, the symbols $\Gamma^{\lambda}{}_{\mu \nu}$ become the Christoffel symbols, which we will denote as $\widehat{\Gamma}^{\lambda}{}_{\mu \nu}\,.$ {The contorsion field $\Lambda\,,$ when contracted with the tetrad in the same way, becomes the antisymmetric part of $\Gamma$\,, which is known as torsion.  }

Using Eq.~(\ref{tetrad-postulate}), the Ricci scalar can be written in terms of the field strength of the spin connection, 
\begin{align}\label{Ricci}
	R(\Gamma) &= F_{\mu\nu}{}^{ab} e^\mu_a e^\nu_b\,, \qquad \qquad {\mathrm{ where}} \\
	F_{\mu\nu}{}^{ab} &= \partial_\mu A_\nu{}^{ab} - \partial_\nu A_\mu{}^{ab} + A_{\mu}{}^{a}{}_{c} A_{\nu}{}^{cb} -  A_{\nu}{}^{a}{}_{c} A_{\mu}{}^{cb}\,.
\end{align}
Then the action of gravity plus a fermion field can be written as 

\begin{equation}\label{action.1}
	S = \int |e| d^4x \left(\frac{1}{2\kappa} F_{\mu\nu}{}^{ab} e^\mu_a e^\nu_b + {\mathscr L}_\psi\right) \,,
\end{equation}
with $\kappa = 8\pi G\,,$ and
\begin{align}
	{\mathscr L}_\psi &= \frac{i}{2}	\left(\bar{\psi}\gamma^\mu\partial_\mu\psi - 
	\partial_\mu\bar{\psi}\gamma^\mu\psi - \frac{i}{4} A_{\mu}{}^{ab} \,
	\bar{\psi}[\sigma_{ab}, \gamma_c ]_{_+} \psi\, e^{\mu c} \right) \notag \\ 
	&\qquad \qquad +im\bar\psi\psi\,.\label{L_psi}
\end{align}
Using the split connection, we can write 
\begin{align}
	S =&  \frac{1}{2\kappa}\int |e| d^4x\, \left(R(\widehat{\Gamma}) +  e^\mu_a e^\nu_b \partial_{[\mu}\Lambda_{\nu]}{}^{ab} 
	+	 e^\mu_a e^\nu_b\left[\omega_{[\mu}, \Lambda_{\nu]}{}^{ab}\right]_{-}\right) \notag \\
	 &   + \int |e| d^4x\, 
	\left(	\frac{1}{2\kappa}  e^\mu_a e^\nu_b\left[\Lambda_{[\mu }, \Lambda_{\nu]}{}^{ab}\right]_{-}  
	+	{\mathscr L}_\psi \right)\,.
\label{action}
\end{align}

The $\Lambda$-terms in the first line contribute a total derivative, so varying $\Lambda$ produces
\begin{equation}\label{spin-connection}
	\Lambda_\mu{}^{ab} = \frac{\kappa}{8}e^c_\mu\,\bar{\psi}[\gamma_c,\sigma^{ab} ]_{+}\psi \,.
\end{equation}
This is totally antisymmetric in $a,b,c$\, because of the identity  $[\gamma_c,\sigma_{ab} ]_{_+} = 2\epsilon_{abcd}\gamma^d\gamma^5 $\,.  Since $\Lambda$ is fully expressible in terms of the other fields without derivatives, it can be replaced in the action by this solution. Then varying the tetrads we get the Einstein equations 
\begin{equation}\label{EE2}
	\widehat{R}_{\mu\nu} - \frac12 g_{\mu\nu} \widehat{R} = \kappa \widehat{T}_{\mu\nu} ({\psi,\bar{\psi}}) -\frac{3 \kappa^2}{16} g_{\mu\nu} \left(\bar{\psi}\gamma_a\gamma_5 \psi \right)^2\,,
\end{equation}
where the hat indicates that the quantity has been calculated with the torsion-free connection. 
The Dirac equation also becomes nonlinear,
\begin{equation}\label{DE1}
	i\slashed{\partial} \psi + \frac{1}{4}\omega_\mu{}^{ab} \gamma_\mu \sigma_{ab} \psi
	+ i m\psi 		- \frac{3\kappa}{8}\left(\bar{\psi}\gamma^a\gamma^5\psi\right)\gamma_a\gamma^5\psi = 0\,.
\end{equation}
The Dirac equation is thus naturally nonlinear in curved spacetime, i.e. whenever $\kappa$ is not assumed to vanish~\cite{Gursey:1957, Finkelstein:1960, Hehl:1971qi, Gasperini:2013}. 

These equations are written for one species of fermions. In general, we need to include all fermion species in the action. Further, since the terms containing $\Lambda$ are invariant on their own, there is no reason to assume that $\Lambda$ will couple identically to all fermion species. For example, if there is even one massless, left-handed neutrino, $\Lambda$ can still couple to it. Thus $\Lambda$ will in general couple to left- and right-handed components of a fermion with different coupling constants. 

Then we can write the generic form of the fermion Lagrangian as
\begin{align}
	{\mathscr L}_\psi &= \sum\limits_{i}	\left(\frac{i}{2}\bar{\psi}_i\gamma^\mu\partial_\mu\psi_i - 
	\frac{i}{2}\partial_\mu\bar{\psi}_i\gamma^\mu\psi_i 	+ \frac{1}{8} \omega_{\mu}{}^{ab} e^{\mu c}  \,
	\bar{\psi}_i[\sigma_{ab}, \gamma_c ]_{_+} \psi_i\, +im\bar\psi_i\psi_i\, \right. \notag \\ 
	&\qquad \qquad \left.
	+ \frac{1}{8} \Lambda_{\mu}{}^{ab} e^{\mu c}
	\left(\lambda^i_{L}\bar{\psi}_{iL} \left[\gamma_c, \sigma_{ab}\right]_{+}\psi_{iL} + \lambda^i_{R}\bar{\psi}_{iR} \left[\gamma_c, \sigma_{ab}\right]_{+}\psi_{iR}\right)
	\right)\,,\label{L_psi_all}
\end{align}
where the sum runs over all species of fermions. The other terms in the action~(\ref{action}) remain as before, so
\begin{equation}\label{chiral.torsion}
	\Lambda_{\mu}{}^{ab} = \frac{\kappa}{4}\epsilon^{abcd}e_{c\mu} \sum\limits_i \left(-\lambda^i_{L}\bar{\psi}_{iL}\gamma_d \psi_{iL} + \lambda^i_{R}\bar{\psi}_{iR}\gamma^d \psi_{iR}\right)\,.
\end{equation}
Since this is totally antisymmetric, the geodesic equation is unaffected and all particles fall at the same rate. 

As before, we can insert this solution back into the action and get an effective quartic interaction term
\begin{equation}\label{4fermi}
	-\frac{3\kappa}{16}\left(\sum\limits_i \left(-\lambda^i_{L}\bar{\psi}^i_{L} \gamma_a \psi^i_L + \lambda^i_{R}\bar{\psi}^i_{R} \gamma_a \psi^i_{R}\right)\right)^2\,.
\end{equation}
We will use the words \textit{torsional} and \textit{geometrical} interchangeably when referring to this term and the associated coupling constants, 
even though torsion itself has disappeared from the action. 
This interaction term is formally independent of the background metric, but of course it affects the metric through Einstein equations. However, the resulting curvature will in general be small enough that the background can be taken to be approximately flat for the purpose of quantum field theory calculations. We emphasize that this term is not an extension of General Relativity, nor any special effect of spacetime curvature on fermions, but how ordinary fermions behave due to the fact that spacetime is not flat. 

The geometrical interaction will contribute to the effective mass of fermions propagating through matter. To see this, we first note that the interaction can be rewritten in terms of vector and axial currents as
\begin{equation}\label{V-A}
	-\frac{1}{2}\left(\sum\limits_i \left(\lambda_i^V \bar{\psi}^i \gamma_a \psi^i + \lambda_i^A \bar{\psi}^i \gamma_a\gamma^5 \psi^i\right)\right)^2\,,
\end{equation}
where we have written $\lambda^{V, A} = \frac{1}{2}(\lambda_R \pm \lambda_L)$\, and absorbed a factor of $\sqrt{\frac{3\kappa}{8}}$\, in $\lambda_{V, A}\,.$
The Dirac equation is thus
\begin{equation}\label{NLD}
	\gamma^\mu  \partial_\mu \psi^i - \frac{i}{4}\omega_{\mu}{}^{ab} \gamma^\mu \sigma_{ab} \psi^i
	+  m\psi^i
	+i\left(\sum_f\left(\lambda^f_{V}\bar{\psi}^f \gamma_a \psi^f + \lambda^f_A \bar{\psi}^f \gamma_a\gamma^5 \psi^f\right)\right)
	\left(\lambda^i_{V}\gamma_a \psi^i + \lambda^i_A \gamma_a\gamma^5 \psi^i\right) = 0\,,
\end{equation}
where $\psi^i$ is the fermion field whose equation we want, and $\psi^f$ is the field of any fermion, the sum running over all species as before. For a fermion passing through background fermionic matter at ordinary densities (i.e. not producing a large curvature), with no other source of curvature being present, we can ignore $\omega_\mu{}^{ab}$\,. Also, since the interaction is quite weak, we can replace the term in the parentheses other than for $f=i\,$ by its average value,
\begin{equation}\label{avg}
	\sum_{f\neq i}\left\langle\lambda^f_{V}\bar{\psi}^f \gamma_a \psi^f + \lambda^f_A \bar{\psi}^f \gamma_a\gamma^5 \psi^f\right\rangle\,.
\end{equation}
If the fermion is passing through ordinary matter, the fermions $f$ are electrons, protons, and neutrons (or electrons and $u$ and $d$ quarks). If this background matter is at rest on average in some frame, the space components of the axial current average to the spin density, while the 0-th component gives the axial charge. Both of these are approximately zero except is special cases. The space components of the first term average to momentum density, which also vanishes. Thus we are left with only the 0-th component of the first term, which averages to the number density of the fermions of type $f$\,. We can therefore write the Dirac equation for the $i$-type fermion  as 
\begin{align}
	&	\gamma^\mu  \partial_\mu \psi^i - \frac{i}{4}\omega_\mu{}^{ab} \gamma^\mu \sigma_{ab} \psi^i
	+  m\psi^i + i\tilde{n}^i_\lambda \left(\lambda^i_V \gamma_0\psi^i + \lambda^i_A \gamma_0\gamma^5 \psi^i\right)\, = 0\,,
	\label{NLD2}
\end{align}
where $\tilde{n}^i_\lambda$ is related to the density of the background fermions as $\tilde{n}^i_\lambda = \displaystyle\sum_{f\neq i}\lambda^f_{V}\left\langle{\psi}^{f\dagger} \psi^f\right\rangle\,,$ and we have suppressed all gauge interactions. We have also ignored self-interactions, which can be done if the $i$-type fermion in this equation is not electron, proton or neutron. We note that the matter effect in this equation can also be derived in a completely covariant manner using thermal field theory~\cite{Pal:1989xs,Notzold:1987ik, Ghose:2023ttq}.
The coupling-weighted density $\tilde{n}$ contributes to the effective mass of fermions in a matter background. This can potentially affect models of many physical phenomena which concern the passage of particles through matter. Stellar models involve interactions of fermions with background gas made of fermions; dark matter
estimates require the estimation of mass-luminosity relations, which would be modified if the effective mass of baryons in dense stars is significantly different from that in normal stars; models of early universe cosmology can also be affected for the same reason. The torsional four-fermion interaction will also contribute
to all fermion-fermion scattering processes, e.g. in accelerator experiments, although the contribution must be sufficiently small so as to have been masked by other interactions at current energies.

In this paper, we will be concerned with how neutrino oscillations are affected by this geometrical four-fermion interaction. {It is well known that neutrinos passing through normal matter receive a contribution to their Hamiltonian due to weak interactions with the background~\cite{Wolfenstein:1977ue, Mikheev:1986gs, Mikheev:1986wj}.} While it is believed that even in regions of strong gravity such as supernovae, matter effects will wash out the effect of curvature on neutrino oscillations~\cite{Cardall:1996cd}, what we have
here is different from ordinary gravitational effects. That is because the geometrical interaction is not affected by the local curvature of spacetime, and also because the dimensionful coupling constants $\lambda_{V, A}$ cannot be determined from theoretical considerations, but must be fixed by experiments. 
We will come back to this point in the last section, for the moment let us be agnostic about the size of these coupling constants.  

{In the absence of flavor mixing,} fermions in a weak doublet should have the same coupling to torsion. For example, $u$ and $d$ quarks, and thus protons and neutrons, should have the same $\lambda_L$\,. But $\lambda_R$ can be different for different fermions, so the $\lambda_V$ and $\lambda_A$ in Eq.~(\ref{V-A}) can be  different for different fermions in general. 

\section{Neutrino mixing and oscillation: Two neutrinos}\label{2nu}
Let us work with two species of neutrinos to begin with, passing through normal matter at uniform density. It is the fields in the mass basis which couple to torsion, since torsion appears as part of the geometric connection. The interaction Lagrangian is\footnote{In this section and below we will use the signature $(+---)$ as that is more common in particle physics.}
\begin{equation}\label{L-nu.1}
		-\left(\sum_{i=1,2}\left(\lambda_{i}^{V}\bar{\nu}_i \gamma_a \nu_i + \lambda_{i}^A \bar{\nu}_i \gamma_a\gamma^5 \nu_i \right)\right) \times
			\left(\sum_{f=e, p, n}\left(\lambda_{f}^{V}\bar{f} \gamma_a f + \lambda_{f}^A \bar{f} \gamma_a\gamma^5 f \right)\right) 
		\,.
\end{equation}
We have ignored neutrinos in the second factor above since their background density in normal matter is several orders of magnitude smaller than those of electrons or baryons. We also note that except in very special cases, the density of matter is not sufficient to cause high curvature -- even neutron stars have densities of the order of $10^{-80}$ in Planck units. Therefore, in most physical situations, $\omega_\mu{}^{ab}$ can be set to zero and the calculations can be done as in a flat background. 

The factor with the background fermions can be replaced by its average value by considering the forward scattering amplitude of neutrinos {in a manner analogous to the case of weak interactions~\cite{Wolfenstein:1977ue}}.
For a background distribution of non-relativistic fermions, the average for the fermions of type $f$ becomes the 
number density $n_f$\,~\cite{Pal:1989xs,Ghose:2023ttq}. Thus the interaction term is
\begin{equation}\label{L-nu.eff}
		-\left(\sum_{i=1,2}\left(\lambda_{i}^{V}\bar{\nu}_i \gamma_0 \nu_i + \lambda_{i}^A \bar{\nu}_i\gamma_0 \gamma^5 \nu_i \right)\right) \, \tilde{n}\,,
\end{equation}
where $\tilde{n}$ is the related to the number density of the background matter as $\tilde{n} = \sum\limits_{f=e,p,n}\lambda_{f} n_f$\,; it is what we called $\tilde{n}^i_\lambda$ earlier, for $i=\nu\,.$
We now assume that for right handed neutrinos, the torsional coupling is negligible compared to that for left handed neutrinos\footnote{In the absence of mixing, fermions in a weak doublet (e.g. $u$ and $d$ quarks) should have the same $\lambda_L$\,. It is not necessary to set $\lambda_R = 0$\,, but we will do so for all fermions in this paper, for convenience. However, the couplings $\lambda_L$ must be different from one {neutrino mass eigenstate} to another if there is to be an effect on neutrino oscillations.} (i.e. maximally chiral).
Then the contribution to the effective Hamiltonian is
\begin{equation}\label{H-nu.eff}
	\sum_{i=1,2}\left(\lambda_{i}{\nu}_i^\dagger\mathbb{L} \nu_i  \right)\,\tilde{n}\,.
\end{equation}

The mass eigenstates $\ket{\nu_i}$ and the flavor eigenstates $\ket{\nu_{\alpha}}$ are related to each other by
\begin{equation}
	\ket{\nu_{\alpha}}=\sum_{i}U^{*}_{\alpha i}\ket{\nu_{i}}\,,
	\label{2.mixing}
\end{equation}
where $U$ is the mixing matrix, $U=\begin{pmatrix}\cos\theta & \sin\theta \\ -\sin\theta\, & \cos\theta\end{pmatrix}$\,.
We can now write the Schr\"{o}dinger equation for the neutrinos in the flavor basis as
\begin{equation}
	i\frac{d}{dt}\begin{pmatrix}{\nu_e} \\ {\nu_{\mu}}\end{pmatrix} =\left[E_0 {\mathbb I} +\frac{1}{4E}\begin{pmatrix}-\Delta m_s^2\cos 2\theta +D & \Delta m_s^2\sin 2\theta\\ \Delta m_s^2 \sin 2\theta & \Delta m_s^2 \cos2\theta - D\end{pmatrix}\right]\begin{pmatrix}{\nu_e} \\ {\nu_{\mu}}\end{pmatrix}\,.
	\label{eq:TDSE_for_2nu_2}
\end{equation}
We have included the contribution from weak interactions with the background, and written   $D = 2\sqrt{2}G_F n_e E\,,$ with
\begin{equation}
	E_0 = E+\frac{m_1^2+m_2^2}{4E}+\frac{\lambda_1+\lambda_2}{2}\tilde{n} -\frac{G_F}{\sqrt{2}}(n_n-n_e)\,,
\end{equation}
and also defined a modified $\Delta m^2$ as
\begin{equation}\label{ms-squared}
		\Delta m_s^2 =\Delta m^2+2 \tilde{n}E \Delta \lambda\,,
\end{equation}
{where} $\Delta m^2=m_2^2-m_1^2\,$ and $\Delta\lambda = \lambda_2 - \lambda_1\,.$
Let us write $\theta_M$ for the mixing angle in matter, modified by the torsional interaction,
\begin{equation}
	\tan 2\theta_M=\frac{\tan 2\theta}{1-\frac{D}{\Delta m_s^2 \cos 2\theta}}\,.
\end{equation}
%
Then we can diagonalize Eq.~(\ref{eq:TDSE_for_2nu_2}), 
 and calculate  the survival probability
\begin{equation}
	P_{\nu_e \to \nu_e}=1-\sin ^2(2\theta_M)\sin^2\left(\frac{\Delta m_M^2}{4E}L\right)\, \label{nu_e-survival-2nu}
\end{equation}
and the conversion probability 
\begin{equation}
	P_{\nu_e \to \nu_{\mu}}=\sin ^2(2\theta_M)\sin^2\left(\frac{\Delta m_M^2}{4E}L\right)\,,
	 \label{nu_-nu_mu-2nu}
\end{equation}
where we have written $\Delta m_M^2=\sqrt{(\Delta m_s^2\cos 2\theta-D)^2+(\Delta m_s^2 \sin 2\theta)^2}\,.$ 
These probabilities have the same form as the usual one. Thus the torsional four-fermion interaction modifies the mass squared difference of neutrinos via Eq.~(\ref{ms-squared}) and may be hidden in the observed values of $\Delta m^2$\,. In particular, the effective mass squared difference in matter is a function of the energy of the incident neutrinos. This is different from the usual theory of neutrino oscillations. 

{ We have mentioned before that} other authors have also considered scenarios of neutrino oscillation with torsion~\cite{DeSabbata:1981ek,  Zhang:2000ue, Adak:2000tp, Fabbri:2015jaa}. However, there are three important differences between these works and the model we have presented here: \textit{a)} we do not consider a background torsion field, but only that which is dynamically generated by the presence of fermions, so it is a purely auxiliary field which can be integrated out; \textit{b)} we consider a chiral coupling of the torsion to fermions, unlike in the existing literature, { and \textit{c)} after integrating out the chiral torsion, we find  a kind of MSW effect, but in the mass basis.} 

\section{Neutrino mixing and oscillation: Three neutrinos}\label{3nu}
Let us now consider the effect of the torsional four-fermion interaction when there are three species of neutrinos. In this case the mixing matrix contains a complex phase for CP-violation~\cite{Kobayashi:1973fv, Cabibbo:1977nk, Walsh:2022pqg, Fuller:2022nbn}. We follow the conventions of the Review of Particle Physics (RPP)~\cite{ParticleDataGroup:2022pth} and write the mixing matrix as
\begin{align}
	U&=\begin{pmatrix}c_{12}c_{13} & s_{12}c_{13} & s_{13}e^{-i\delta}\\-s_{12}c_{23}-c_{12}s_{23}s_{13}e^{i\delta} & c_{12}c_{23}-s_{12}s_{23}s_{13}e^{i\delta} & s_{23}c_{13}\\ s_{12}s_{23}-c_{12}c_{23}s_{13}e^{i\delta} & -c_{12}s_{23}-s_{12}c_{23}s_{13}e^{i\delta} & c_{23}c_{13}\end{pmatrix}\begin{pmatrix}e^{i\eta_1}&0&0\\0&e^{i\eta_2}&0\\0&0&1\end{pmatrix}\,,
	\label{3nu-mixing}
\end{align}
where $c_{ij}=\cos\theta_{ij}$ and $s_{ij}=\sin\theta_{ij}\,.$ In general, the angles $\theta_{ij}$ can be taken to lie in the first quadrant, while the CP-violation phase $\delta$ and the Majorana phase angles are taken to be between $0$ and $2\pi$\,. 
In the case of Dirac neutrinos the phases can be absorbed into the neutrino phases,  so the mixing matrix can be conveniently expressed as a product of rotation matrices ${\mathcal O}_{ij}$ for rotation in the $ij$-plane and $U_\delta = \mathrm{diag}(1,\, 1,\, e^{i\delta})$\,. Then the matrix can be written as
\begin{equation}
	U=\mathcal{O}_{23}\mathcal{U}_{\delta}\mathcal{O}_{13}\mathcal{U}_{\delta}^{\dagger}\mathcal{O}_{12}\,. \label{def:mixing_matrix_product_of}
\end{equation}
We can write the Schr\"odinger equation for the mass eigenstates as~\cite{Barick:2023qjq}
\begin{align}
i\frac{d}{dt}\begin{pmatrix}{\nu_1}\\ {\nu_2} \\ \nu_3\end{pmatrix}=\left[E+\frac{1}{2E}\begin{pmatrix}m_1^2 & 0 & 0 \\ 0 & m_2^2 & 0 \\ 0 & 0 & m_3^2\end{pmatrix}+\begin{pmatrix}\lambda_1 & 0 & 0 \\ 0 & \lambda_2 & 0 \\ 0 & 0 & \lambda_3\end{pmatrix}\tilde{n}-\frac{G_F}{\sqrt{2}}n_n+U^{T}\begin{pmatrix}2A & 0 & 0 \\ 0 & 0 & 0 \\ 0 & 0 & 0\end{pmatrix}U^{*}\right]\begin{pmatrix}\nu_1 \\ \nu_2 \\ \nu_3 \end{pmatrix}\,, \label{eq:TDSE_for_3nu.a}
\end{align}
where $A=\frac{G_F}{\sqrt{2}}n_e$\,. If the matter density is uniform, or the variation is slow enough, the equation for the evolution of flavor eigenstates can be written as\footnote{In~\cite{Akhmedov:2004ny} as well as in the RPP~\cite{ParticleDataGroup:2022pth}, $U$ and $U^\dagger$ appear in place of $U^*$ and $U^T$, respectively. That is possibly a typo, as we start from the same Eq.~(\ref{2.mixing}) and find the same final results when $\lambda_i=0$ as those works.}
\begin{align}
	i\frac{d}{dt}\threeflavor&=\left[E'_0\mathbb{I}+\frac{1}{2E}\left[U^*\begin{pmatrix}0 ~& 0 & 0 \\ 0 ~& \Delta \tilde{m}_{21}^2 & 0 \\ 0 ~& 0 & \Delta \tilde{m}_{31}^2\end{pmatrix}U^T+\begin{pmatrix}D & 0 & 0 \\ 0 & 0 & 0 \\ 0 & 0 & 0\end{pmatrix}\right]\right]\,\threeflavor \,.\label{eq:TDSE_for_3nu.b}
\end{align}
We have used the definition 
\begin{align}
E'_0=E+\frac{m_1^2+2\lambda_1\tilde{n}E}{2E}-\frac{G_F}{\sqrt{2}}n_n\,, 
\end{align}
and $D = 2\sqrt{2}G_F n_e E\,$ as before.
We wish to find the eigenvalues of this matrix by the method of perturbation using a small parameter. For this, let us first define
\begin{equation}\label{tm-squared}
	\Delta \tilde{m}^2_{ij} := \Delta m^2_{ij} +2 \tilde{n} E \Delta\lambda_{ij}\,,
\end{equation}
{where} $\Delta m_{ij}^2=m_i^2-m_j^2$ and $\Delta \lambda_{ij}=\lambda_{i}-\lambda_{j}\,.$ We will generally use a tilde over a symbol to indicate that it includes torsional coupling constants. Then we can rewrite Eq.~(\ref{eq:TDSE_for_3nu.b}) as 
\begin{align}
	i\frac{d}{dt}\threeflavor &=\frac{\Delta \tilde{m}_{31}^2}{2E}\left[U^*\begin{pmatrix}0 & 0 & 0 \\ 0 & \alpha & 0 \\ 0 & 0 & 1\end{pmatrix}U^T+\begin{pmatrix} \tilde{A} & 0 & 0 \\ 0 & 0 & 0 \\ 0 & 0 & 0\end{pmatrix}\right]\threeflavor \nonumber \\
	&= \frac{\Delta \tilde{m}_{31}^2}{2E}{\cal O}_{23}{\cal U}_\delta^{*} M {\cal U}_\delta^{T} {\cal O}_{23}^T\threeflavor\,,\label{eq:TDSE_for_3nu_short} 
\end{align}
where we have suppressed matrices proportional to the identity matrix as those will contribute to a common phase for all the neutrinos and will not contribute to the transition probability.
We have also written
\begin{equation}\label{M.def}
	M={\cal O}_{13}{\cal O}_{12}\begin{pmatrix}0 & 0 & 0\\0 & \alpha & 0\\0 & 0 & 1\end{pmatrix}
	{\cal O}_{12}^T {\cal O}_{13}^T+\begin{pmatrix} \tilde{A} & 0 & 0\\0 & 0 & 0\\0 & 0 & 0\end{pmatrix} \,.
\end{equation}
In these expressions, $\tilde{A} =D/\Delta \tilde{m}_{31}^2 = 2\sqrt{2}G_F n_eE/\Delta \tilde{m}_{31}^2\,,$ a dimensionless quantity.
In order to find the mixing matrix, we first diagonalize $M$ by finding the eigenvalues by the method of perturbation, treating $s_{13}$ and $\alpha= \Delta \tilde{m}_{21}^2/\Delta \tilde{m}_{31}^2 \equiv (\Delta m_{21}^2+2\tilde{n} E\Delta \lambda_{21})/(\Delta m_{31}^2+2\tilde{n} E \Delta \lambda_{31})$\, as small parameters~\cite{Akhmedov:2004ny,Nunokawa:2005nx}. 
The eigenvalues and eigenvectors {of $M$} are estimated up to the second order in $\alpha$ and $s_{13}$ using perturbation theory~\cite{Barick:2023qjq}. The eigenvalues are
\begin{align}
	\mu_1&=\tilde{A}+\alpha s_{12}^2+s_{13}^2\frac{\tilde{A}}{\tilde{A}-1}+\frac{\alpha^2 \sin^2(2\theta_{12})}{4\tilde{A}} \label{eq:first_eigenvalue}\\
	\mu_2&=\alpha c_{12}^2-\frac{\alpha^2 \sin^2(2\theta_{12})}{4\tilde{A}} \label{eq:second_eigenvalue}\\
	\mu_3&=1-s_{13}^2\frac{\tilde{A}}{\tilde{A}-1}\,. \label{eq:third_eigenvalue}
\end{align}
Clearly this analysis works only if $\tilde{A}\neq 0$ or 1. 
The Hamiltonian in Eq.~(\ref{eq:TDSE_for_3nu_short}) is related to $M$ by a unitary transformation. Hence, the energy eigenvalues of the Hamiltonian are 
\begin{equation}\label{energy-ev}
	E_i = \frac{\Delta \tilde{m}_{31}^2}{2E} \mu_i\,.
\end{equation}
If we write $W$ for the matrix which diagonalizes $M$ into $\hat{M}$\,, we can rewrite Eq.~(\ref{eq:TDSE_for_3nu_short}) as
\begin{align}
	i\frac{d}{dx}\threeflavor&=\frac{\Delta \tilde{m}_{31}^2}{2E}{\cal O}_{23}{\cal U}_\delta^{*}W\hat{M}W^T{\cal U}_\delta^T{\cal O}_{23}^T\threeflavor \nonumber \\
	&=\frac{\Delta \tilde{m}_{31}^2}{2E}U'^{*}\hat{M}U'^{T}\threeflavor = \frac{\Delta \tilde{m}_{31}^2}{2E}U'^{*}HU'^{T}\threeflavor\,, \label{eq:TDSE_for_3nu_2}
\end{align}
where $U'$ is the mixing matrix, $U'={\cal O}_{23}{\cal U}_\delta W$\,.
{Neutrino oscillation probabilities have been calculated using some approximations in the literature~\cite{Akhmedov:2004ny,Nunokawa:2005nx,Minakata:2006gq,Barger:1980tf, Zaglauer:1988gz, Yasuda:1998mh, Kimura:2002wd, Freund:2001pn, ParticleDataGroup:2022pth}. In order to compare with the analytical expressions in those works, we will evaluate some neutrino oscillation probabilities by including the effect of the geometrical coupling.}
\subsection{{Terrestrial conversion of $\nu_\mu$}}
Let us consider oscillations of muon neutrinos into other flavors in the Earth's crust which has a matter density $n$ of 2.7 gm/cm$^3$\,, for an energy range 0.5 GeV $< E <$ 10 GeV,  which are appropriate for the DUNE experiment~\cite{DUNE:2022aul,DUNE:2020jqi, Ioannisian:2018qwl}. Also, only for the purpose of estimates, let us take all $\lambda_f$ and $\Delta\lambda_{ij}$ to be of the same order of magnitude, and write $\lambda^2$ to mean all quadratic combinations of the form $\lambda_f\Delta\lambda_{ij}\,.$ Then, {for $\lambda^2\sim 0.1 G_F$, we find that  $\alpha$ is a small parameter of the same order of magnitude as $s_{13}$.
Keeping terms up to the 2nd order in $\alpha$ and $s_{13}\,,$} 
we can write the amplitude of $\nu_\mu \to \nu_\tau$ conversion as\footnote{{For $\lambda^2\sim 1.0G_F\,,$ we cannot treat $\alpha$ as a small parameter, but we can numerically find the eigenvalues and the conversion and survival probabilities without using perturbation theory. We will do this below.}}
\begin{align}
	A_{\mu \tau} =& \bra{\nu_{\tau}}e^{-iHL}\ket{\nu_\mu} \nonumber \\
	=&\left(-\frac{\alpha^2}{\tilde{A}^2}s^2_{12}c^2_{12}s_{23}{c}_{23}+\frac{\alpha {s}_{13}}{\tilde{A}(\tilde{A}-1)}{s}_{12}{c}_{12}\left(c^2_{23}{e^{i \delta}}-s^2_{23}{e^{-i \delta}}\right)+\frac{s^2_{13}}{(\tilde{A}-1)^2}{s}_{23}{c}_{23}\right){e^{-iE_1L}} \nonumber \\
	&+\left(-{s}_{23}{c}_{23}+\frac{\alpha^2}{\tilde{A}^2}s^2_{12}c^2_{12}{s}_{23}{c}_{23}+\alpha {s}_{13} (1+\frac{1}{\tilde{A}}){s}_{12}{c}_{12}\left(c^2_{23}{e^{i \delta}}-s^2_{23}{e^{-i \delta}}\right)\right)e^{-i{E_2}L}\nonumber \\
	&+\left({s}_{23}{c}_{23}-\frac{s^2_{13}}{(\tilde{A}-1)^2}{s}_{23}{c}_{23}+\alpha {s}_{13}\frac{\tilde{A}}{\tilde{A}-1}{s}_{12}{c}_{12}\left(s^2_{23}{e^{-i \delta}}-c^2_{23}{e^{i \delta}}\right)\right)e^{-i{E_3}L}\,.
\end{align}
%
Then the probability is calculated to be (with $E_{ij}= E_i - E_j$)
\begin{align}
P_{\mu\tau}=&2s_{23}^2c_{23}^2(1-\cos(E_{23}L))-2\frac{\alpha^2}{\tilde{A}^2}s_{12}^2c_{12}^2s_{23}^2c_{23}^2(1-\cos(E_{12}L)+\cos(E_{13}L)-\cos(E_{23}L)) \nonumber \\
&-2\frac{s_{13}^2}{(\tilde{A}-1)^2}s_{23}^2c_{23}^2(1+\cos(E_{12}L)-\cos(E_{13}L)-\cos(E_{23}L)) \nonumber \\
&+2\alpha s_{13}s_{12}c_{12}s_{23}^3c_{23}\left[\left(1+\frac{1}{\tilde{A}}\right)(\cos\delta-\cos(E_{23}L+\delta))\right. \nonumber \\
&\left.+\frac{1}{\tilde{A}(\tilde{A}-1)}(\cos(E_{12}L+\delta)-\cos(E_{13}L+\delta))+\frac{\tilde{A}}{\tilde{A}-1}(\cos\delta-\cos(E_{23}L-\delta))\right] \nonumber \\
&+2\alpha s_{13}s_{12}c_{12}s_{23}c_{23}^3\left[\left(1+\frac{1}{\tilde{A}}\right)(-\cos\delta+\cos(E_{23}L-\delta))\right. \nonumber \\
&\left.+\frac{1}{\tilde{A}(\tilde{A}-1)}(-\cos(E_{12}L-\delta)+\cos(E_{13}L-\delta))+\frac{\tilde{A}}{\tilde{A}-1}(-\cos\delta+\cos(E_{23}L+\delta))\right]\,.
\label{mutau1}
\end{align}
Similarly, for the conversion of $\nu_\mu \to \nu_e$\,, we have
\begin{align}
	A_{\mu e} = & \bra{\nu_{e}}e^{-iHL}\ket{\nu_\mu} \nonumber \\
	= &\left(\frac{\alpha}{\tilde{A}}{s}_{12}{c}_{12}(1+\frac{\alpha}{\tilde{A}}\cos(2\theta_{12})){c}_{23}+\frac{{s}_{13}}{\tilde{A}-1}(1-\frac{\alpha \tilde{A}}{\tilde{A}-1}{s}_{12}^2)e^{-i\delta}{s}_{23}\right)e^{-iE_1L} \nonumber \\
	&+\left(-\frac{\alpha}{\tilde{A}}{s}_{12}{c}_{12}(1+\frac{\alpha}{\tilde{A}}\cos(2\theta_{12})){c}_{23}\right)e^{-iE_2L}+\left(-\frac{{s}_{13}}{\tilde{A}-1}(1-\frac{\alpha}{\tilde{A}-1}s_{12}^2)e^{-i\delta}{s}_{23}\right)e^{-iE_3L}\,.
\end{align}
with the corresponding probability being
\begin{align}
P({\nu_\mu \to \nu_e})=&\frac{\alpha^2}{2\tilde{A}^2} \sin^2(2\theta_{12}){c}_{23}^2 \left( 1-\cos({E_{12}L}) \right) +2 \frac{{s}_{13}^2}{(\tilde{A}-1)^2} {s}_{23}^{2}\left( 1-\cos({E_{13}L}) \right) \nonumber \\
	&+\frac{1}{\tilde{A}(\tilde{A}-1)}2\alpha {s}_{13}s_{12}c_{12}s_{23}c_{23}(\cos\delta - \cos(E_{12}L+\delta) - \cos(E_{13}L-\delta) + \cos(E_{23}L-\delta)) \,.	
	 \label{eq:nu_mu_to_nu_e}
\end{align}
The probability of survival for muon neutrinos is thus 
\begin{align}
P({\nu_\mu\to\nu_\mu})=&1-2c_{23}^2s_{23}^2(1-\cos(E_{23}L)) \nonumber \\
&+2\frac{\alpha^2}{\tilde{A}^2}s_{12}^2c_{12}^2c_{23}^2\left[c_{23}^2\cos(E_{12}L)+s_{23}^2\cos(E_{13}L)-c_{23}^2-s_{23}^2\cos(E_{23}L)\right]\nonumber \\
&+2\frac{s_{13}^2}{(\tilde{A}-1)^2}s_{23}^2\left[c_{23}^2\cos(E_{12}L)+s_{23}^2\cos(E_{13}L)-c_{23}^2\cos(E_{23}L)-s_{23}^2\right] \nonumber \\
&+4\alpha s_{13}s_{12}c_{12}s_{23}c_{23}\cos\delta\left[c_{23}^2\frac{1}{\tilde{A}(\tilde{A}-1)}\cos(E_{12}L)+s_{23}^2\frac{1}{\tilde{A}(\tilde{A}-1)}\cos(E_{13}L)\right. \nonumber \\
&\left.+\left(1+\frac{1}{\tilde{A}}\right)c_{23}^2-c_{23}^2\frac{\tilde{A}}{\tilde{A}-1}\cos(E_{23}L)+s_{23}^2\left(1+\frac{1}{\tilde{A}} \right)\cos(E_{23}L)-\frac{\tilde{A}}{\tilde{A}-1}s_{23}^2\right]\,.
\label{eq:nu_mu_to_nu_mu}
\end{align}
Our expressions agree with those of~\cite{Akhmedov:2004ny,Nunokawa:2005nx,Minakata:2006gq,Barger:1980tf, Zaglauer:1988gz, Yasuda:1998mh, Kimura:2002wd, Freund:2001pn}  in the energy range $0.5~\text{GeV}<E<10~\text{GeV}$\,, when we set the geometrical couplings to zero. The approximations used are $s_{13}\ll 1$\, and $\alpha\ll 1$\,, {keeping terms up to second order}, as in the works cited above. {We note that these approximations are valid when the parameters have the currently known values (quoted below), with $\lambda^2 \sim 0.1 G_F\,.$ }

Of course the probabilities can be calculated numerically, even when the approximations do not hold, directly by solving Eq.~(\ref{eq:TDSE_for_3nu_2}). 
\begin{figure}[thbp]
	\includegraphics[width=7cm]{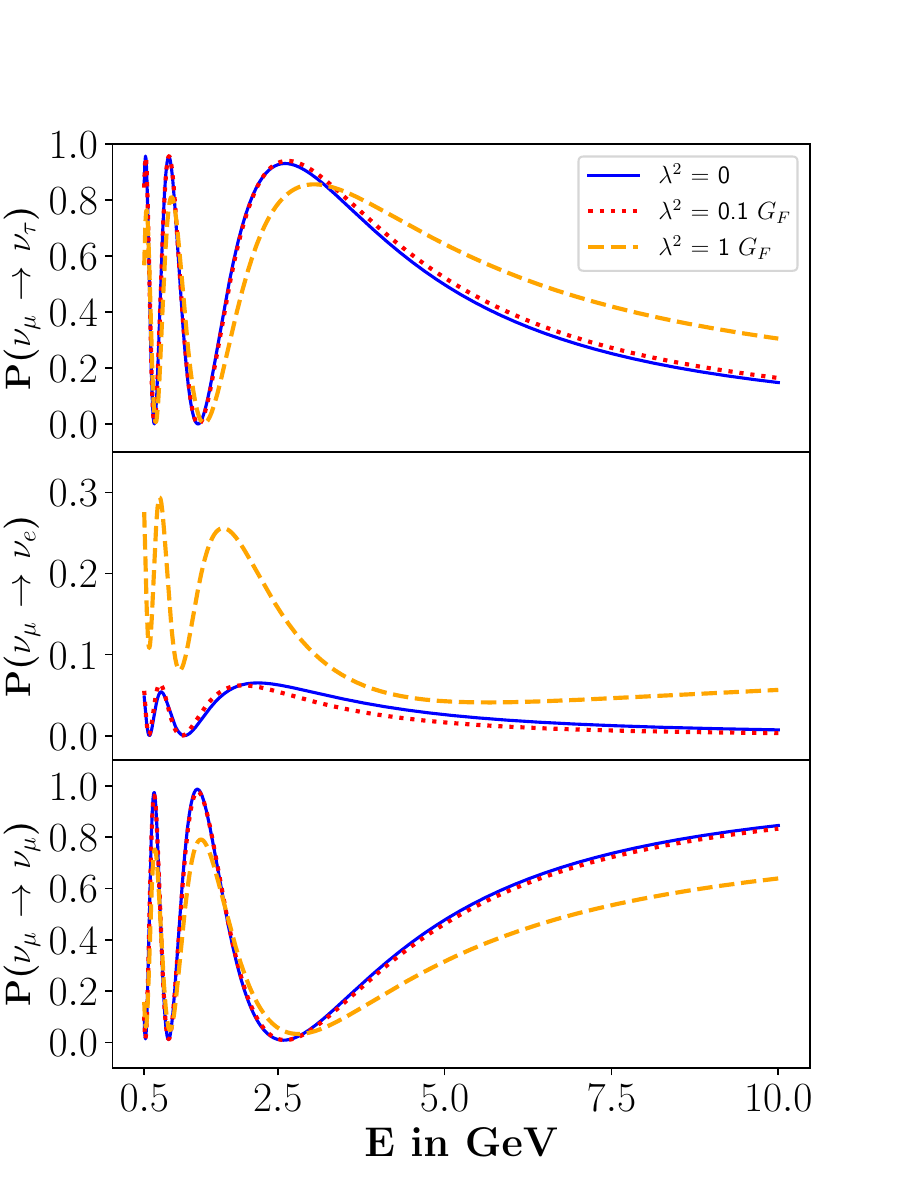}
	\caption{
		$\nu_\mu$ conversion and survival probabilities vs. $E$\,,
		for a baseline $L=1300\,km\,$. $\lambda^2$ denotes the order of magnitude of all quantities quadratic in any $\lambda_f$ and $\Delta\lambda_{ij}\,.$ 
		}
	\centering
	\label{mu-tau-e}
\end{figure}
As a demonstration, let us plot the conversion probabilities for neutrino oscillations in the DUNE experiment with and without the geometrical four-fermion interaction. {While we do not know the size of the couplings $\lambda_i$\,, we can guess that they are not very large compared to $\sqrt{G_F}$\,, since otherwise there would be measurable differences with known results. Some guidance regarding the size of the couplings is provided by neutral current NSI bounds calculated in the literature, assuming that those take into account all possible types of four-fermion interactions.  } 
Let us plot the conversion probabilities when all the $\lambda_f$ are {such that $O(\lambda_f^2) = 1 G_F$ or less} and compare with the case when they are set to zero. 
We have plotted the conversion probabilities
$P(\nu_\mu \to \nu_\tau)\,, P(\nu_\mu\to \nu_e)\,,$ and the survival probability $P(\nu_\mu\to\nu_\mu)\,,$ in Fig.~\ref{mu-tau-e}, 
{by numerically solving the Schr\"odinger equation}. We have used the values~\cite{ParticleDataGroup:2022pth}  
{$\Delta m_{21}^2 = 7.39\times 10^{-23}~\text{GeV}^2\,,$ $\Delta m_{32}^2 = {{2.454\times 10^{-21}~\text{GeV}^2}}\,, \sin^2\theta_{23}=0.563\,, 
\sin^2\theta_{12} = 0.31\,, \sin^2\theta_{13} = 0.02237\,, \delta=221^\circ\,,$} and the average density of the Earth = 2.7 {g/cm$^3$}\,. 
{We have also assumed that $\lambda$ for the electron, proton, and neutron are all of the same order of magnitude so that we can 
write $\tilde{n} \simeq \lambda n\,.$ The oscillation probabilities will then depend on quadratic combinations like 
$\lambda_e\Delta\lambda_{ij}$\,, which we may write as $\lambda\Delta\lambda$ or as $\Delta\lambda^2$ for this purpose. 
As mentioned earlier, we have written $\lambda^2$ to represent the order of magnitude of all quantities quadratic in 
any $\lambda_f$ and $\Delta\lambda_{ij}$\,. In the plots, this  magnitude is taken to be $\sim 0.1 G_F\,$ and 
$\sim 1 G_F\,,$ and we have plotted them along with the case $\lambda=0\,.$} The baseline $L$ is taken to be $1300~km$  
and we have used normal mass ordering. {We see from these plots that if the torsional coupling constants are as large as 
$\sqrt{G_F}\,,$ their effect on the conversion probabilities becomes quite large as the energy grows. 
}

{\subsection{Dependence on the CP angle}
{The weak neutral current interaction is CP-invariant in the absence of mixing --- CP-violation for neutrinos is parametrized in the mixing matrix by the CP phase $\delta_{CP}\,.$ 
Since the geometrical interaction itself causes mixing, i.e. contributes to the off-diagonal terms in the mixing matrix, we can expect that its presence will affect how the conversion probabilities depend on $\delta_{CP}\,.$}

\begin{figure}[htbp]
		\includegraphics[width=5cm]{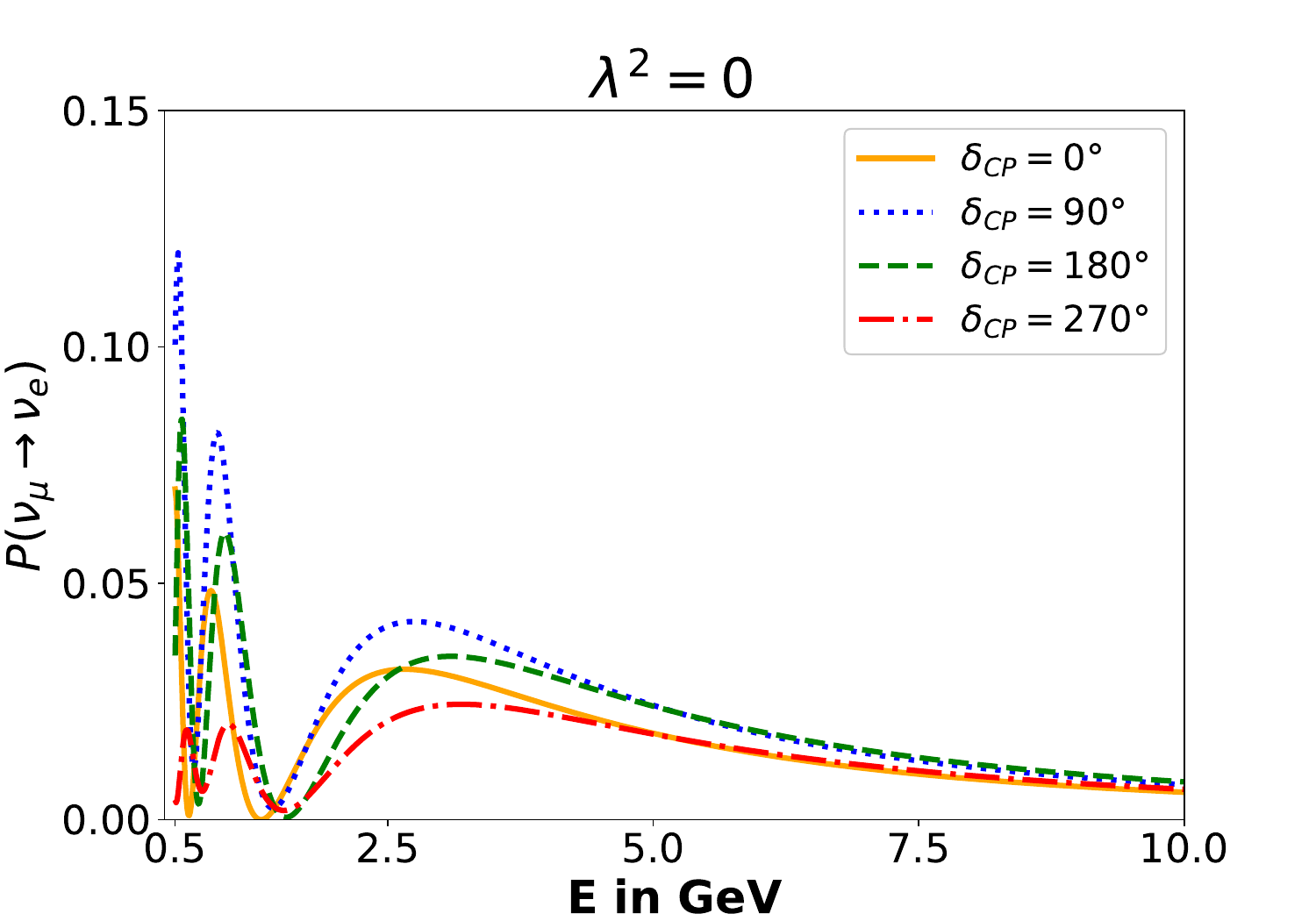}
		\includegraphics[width=5cm]{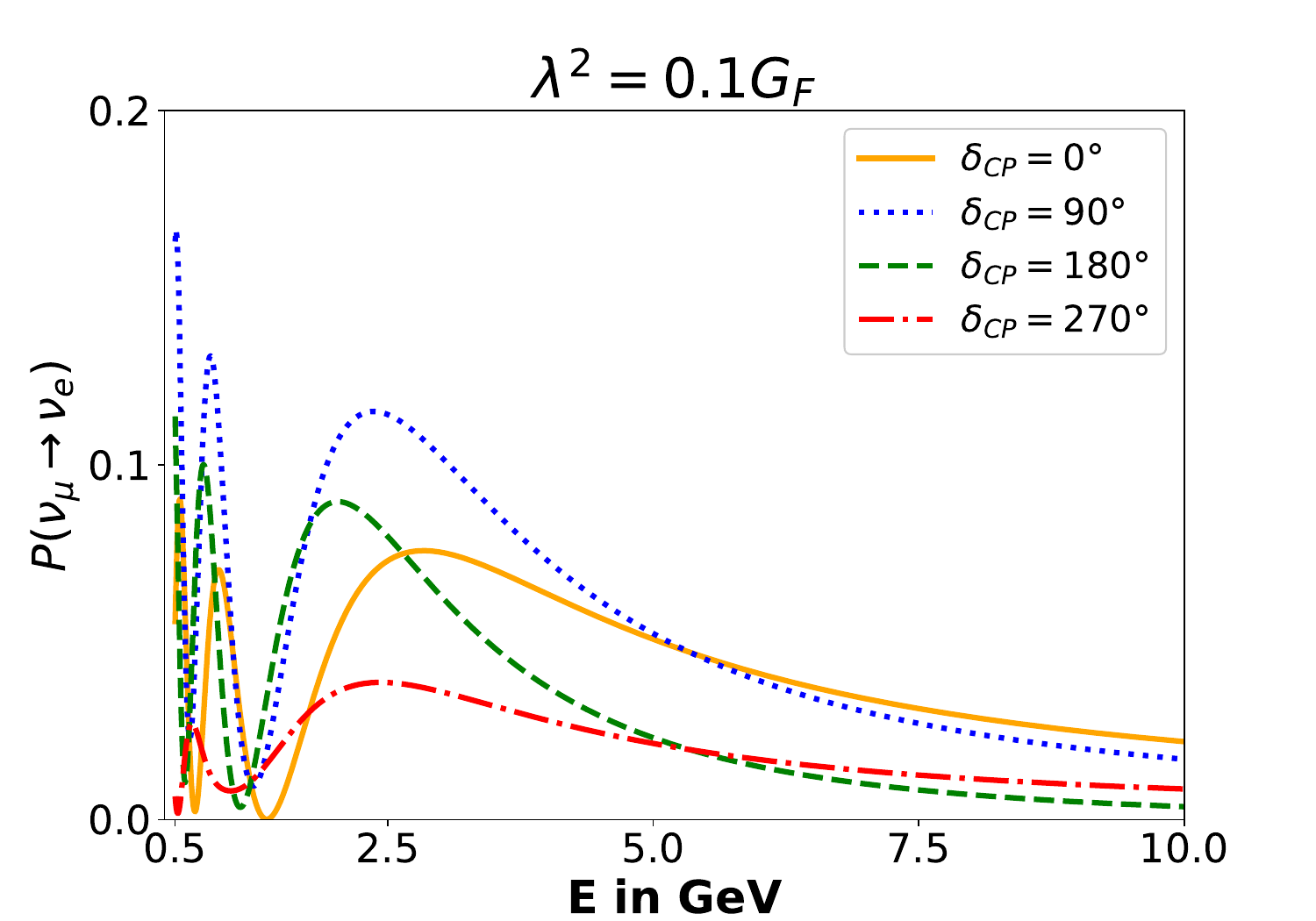}
		\includegraphics[width=5cm]{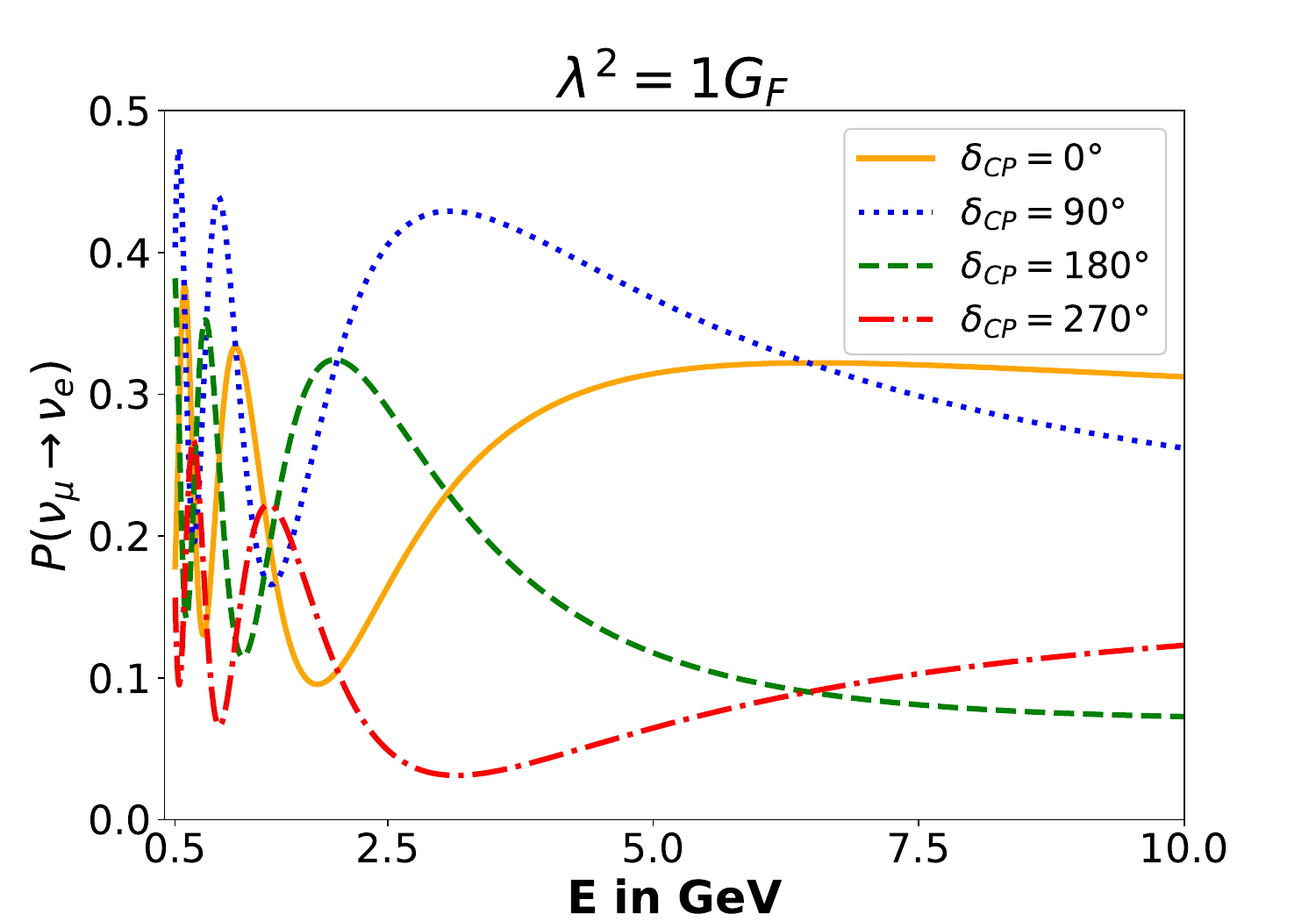}
		\caption{$P_{\mu e}$ vs E for four different values of $\delta_{CP}$ at a baseline of 1300 km and for three values of the geometrical coupling, $\lambda=0$, $\lambda^2=0.1G_F$ and $\lambda^2=1G_F$.}
		\label{mu-e-cp}
\end{figure}

\begin{figure}[htbp]
		\includegraphics[width=5cm]{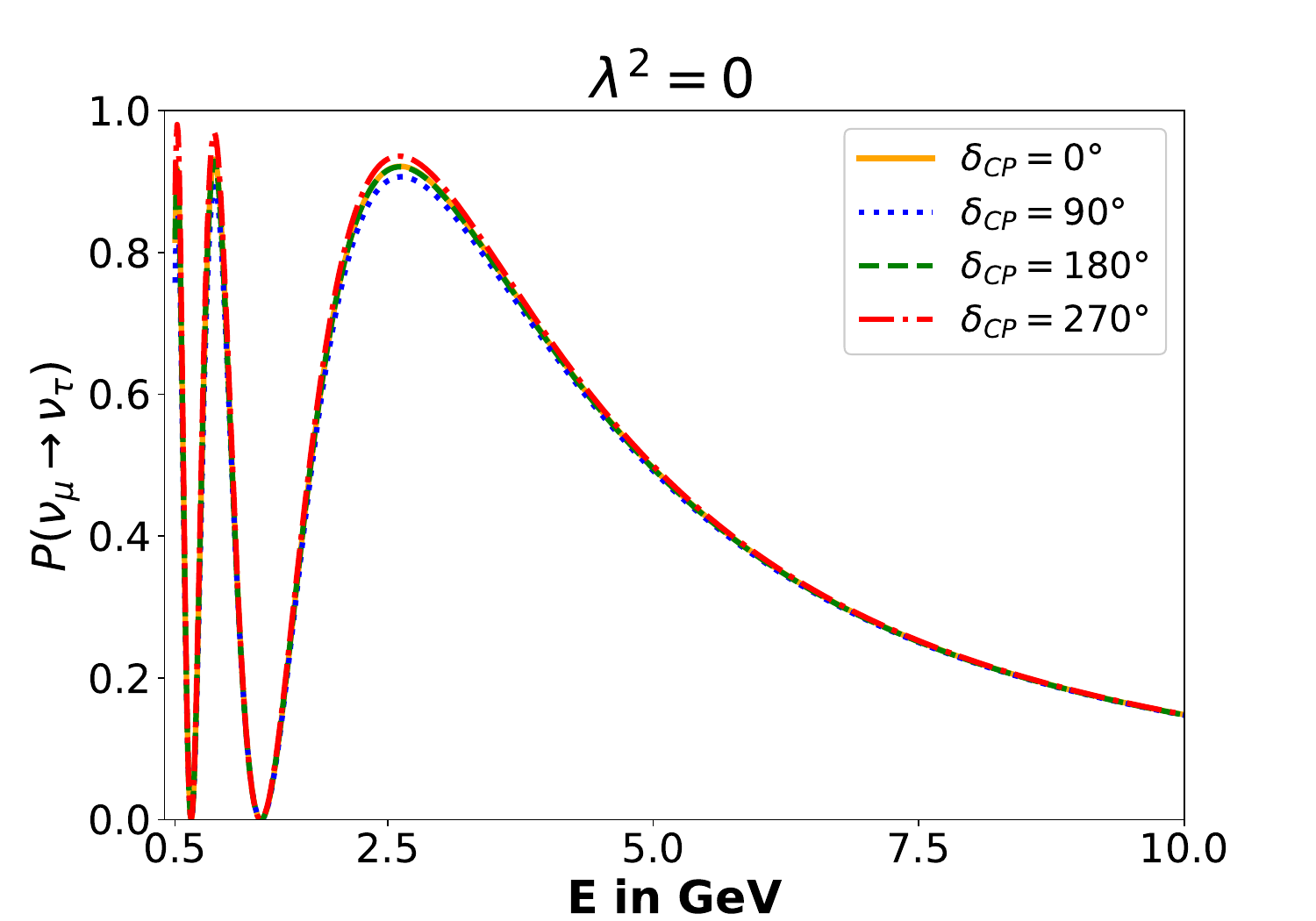}
		\vspace{0.5cm}
		\includegraphics[width=5cm]{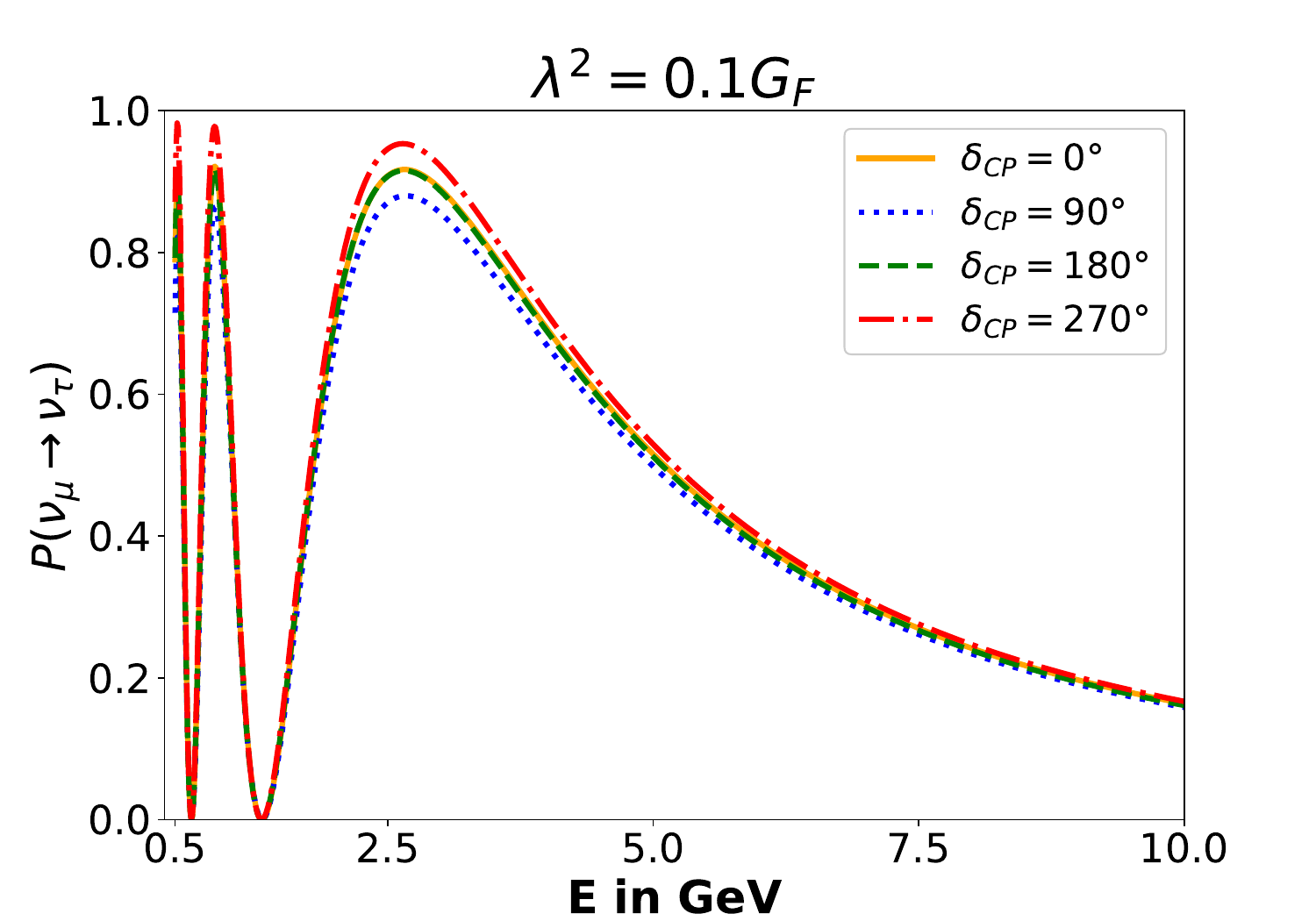}
		\vspace{0.5cm}
		\includegraphics[width=5cm]{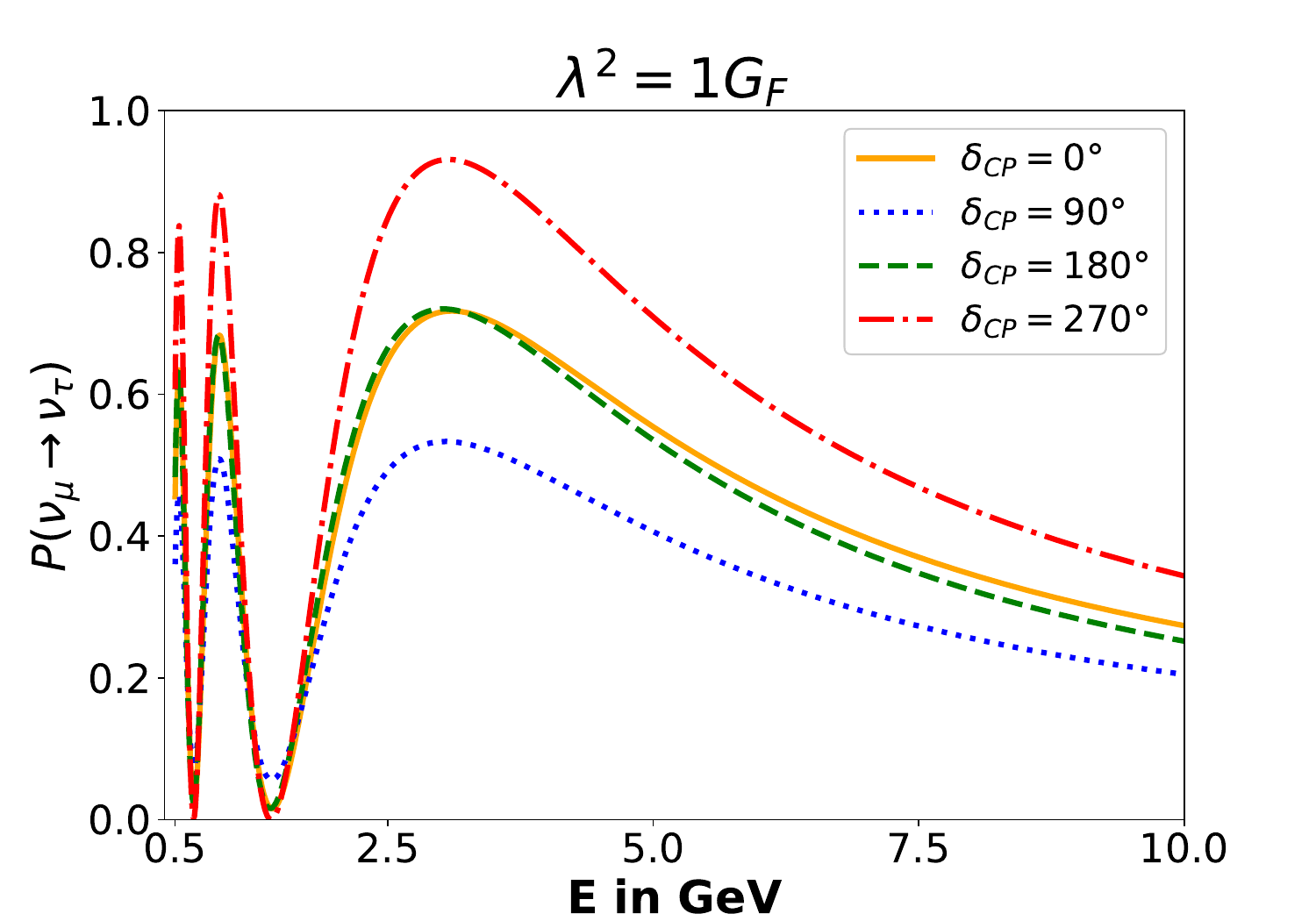}
		\caption{$P_{\mu \tau}$ as a function of E for four different values of $\delta_{CP}$ at a baseline of 1300 km and for three values of the geometrical coupling, $\lambda=0$, $\lambda^2=0.1G_F$ and $\lambda^2=1G_F$.}
		\label{mu-tau-cp}
\end{figure}

\begin{figure}[htbp]
		\includegraphics[width=5cm]{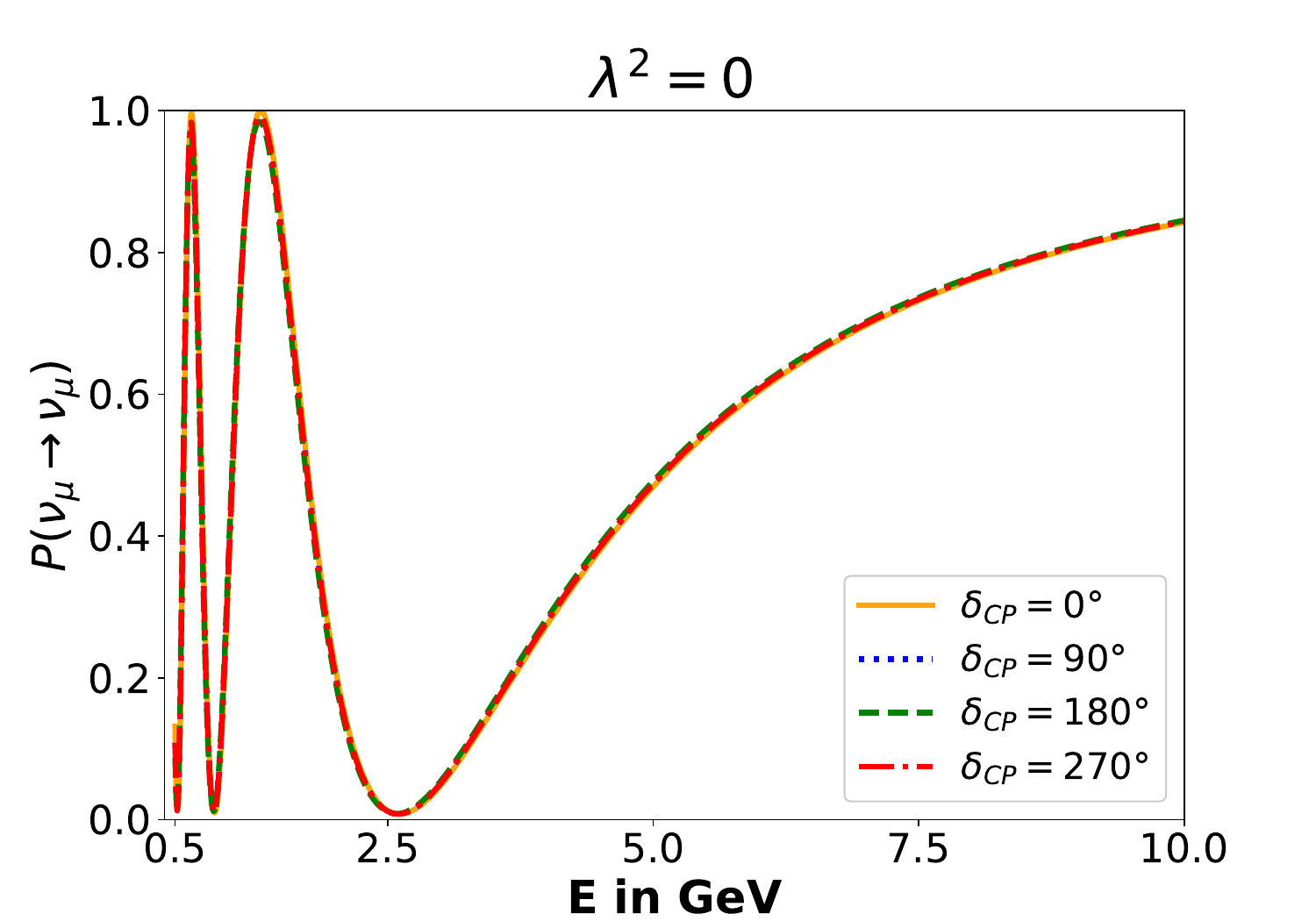}
		\vspace{0.5cm}
		\includegraphics[width=5cm]{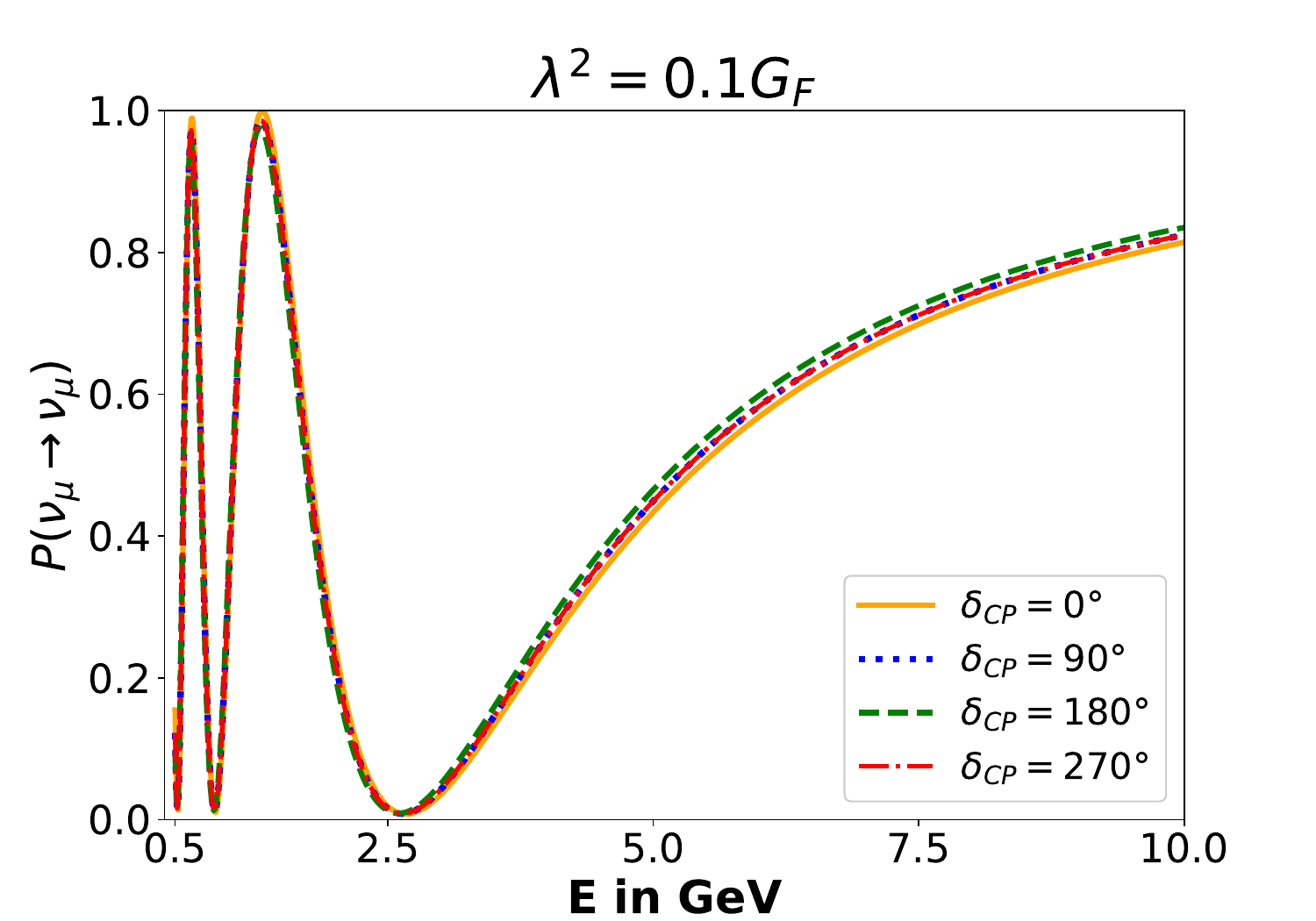}
		\vspace{0.5cm}
		\includegraphics[width=5cm]{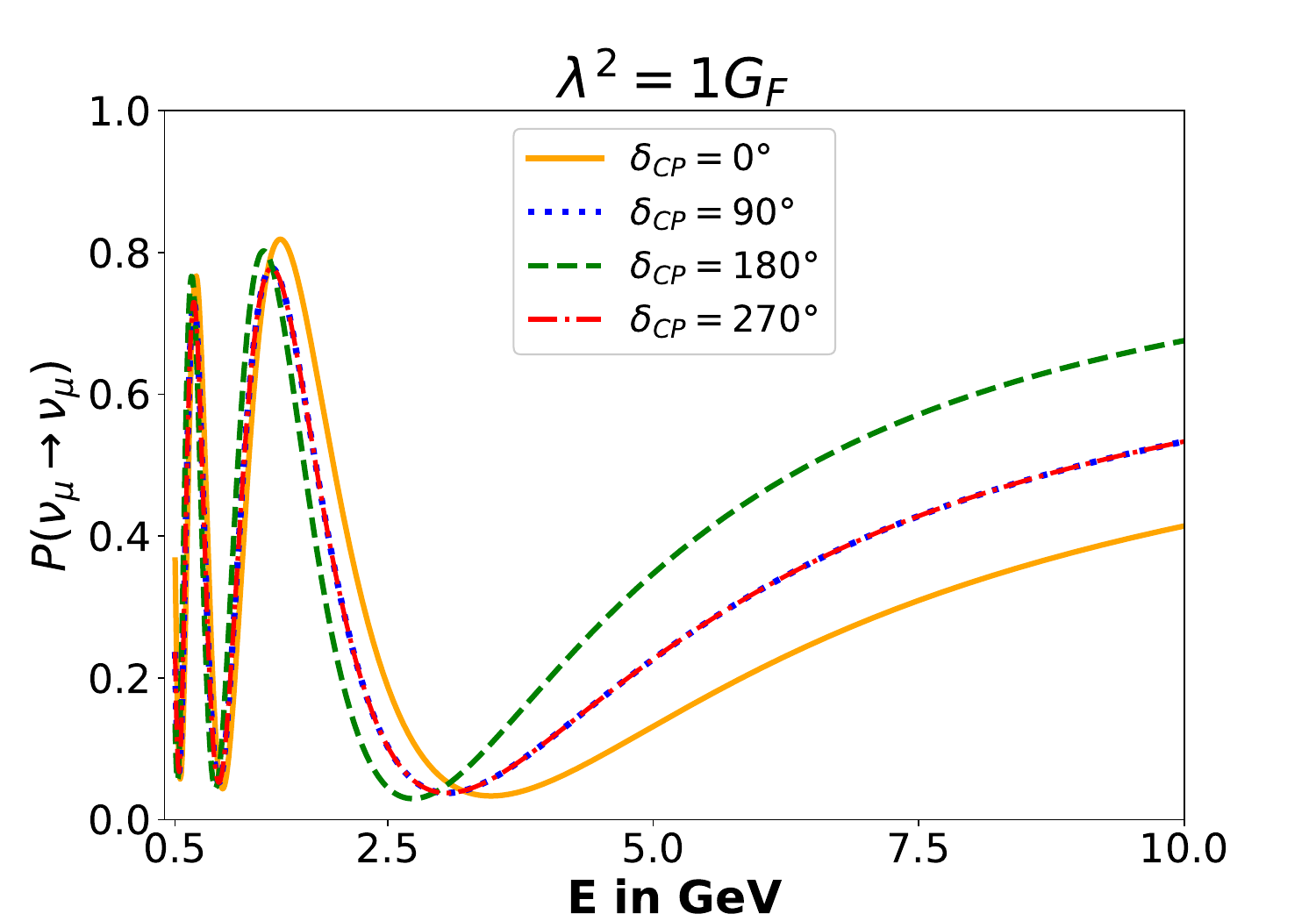}
		\caption{$P_{\mu \mu}$ as a function of E for four different values of $\delta_{CP}$ at a baseline of 1300 km and for three values of the geometrical coupling, $\lambda=0$, $\lambda^2=0.1G_F$ and $\lambda^2=1G_F$.}
		\label{mu-mu-cp}
\end{figure}
{We have therefore plotted the conversion probabilities $P_{\mu e}$ and $P_{\mu \tau}$\,, as well as the survival probability $P_{\mu\mu}$\,, for different values of $\delta_{CP}$ when the geometrical interaction is set to zero and also for two different scales of the geometrical coupling constant.} Here we have used normal mass ordering and the other relevant parameters have been taken from~\cite{ParticleDataGroup:2022pth} including the Super Kamiokande atmospheric data. {It is known that in the absence of geometrical coupling the conversion probabilities depend on the CP phase angle in the low energy region~{\cite{Masud:2015xva, Medhi:2021wxj}}. From the plots in Fig.~\ref{mu-e-cp}, Fig.~\ref{mu-tau-cp}, and Fig.~\ref{mu-mu-cp}, it can be clearly seen that if we turn on the torsional interaction, this dependence on the CP phase angle increases significantly with the increase in the strength of geometrical coupling. }

{\subsection{Terrestrial conversion and survival of $\nu_e, \bar{\nu}_e$}
{Probabilities for the conversion and survival of $\nu_e$ in similar approximations (i.e. terrestrial scenarios) are also found in the literature. We have calculated them in the presence of the geometrical interaction from Eq.~(\ref{eq:TDSE_for_3nu_2}). 
Let us mention them here for the sake of completeness and also for comparison with standard results. }

The $\nu_e \to \nu_\mu$ conversion probability is given by (see~\cite{Barick:2023qjq} for the amplitude)
\begin{align}
P({\nu_e \to \nu_\mu})=&\frac{\alpha^2}{2\tilde{A}^2} \sin^2(2\theta_{12}){c}_{23}^2 \left( 1-\cos({E_{12}L}) \right) +2 \frac{{s}_{13}^2}{(\tilde{A}-1)^2} {s}_{23}^{2}\left( 1-\cos({E_{13}L}) \right) \nonumber \\
	&+\frac{2\alpha }{\tilde{A}(\tilde{A}-1)}{s}_{13}s_{12}c_{12}s_{23}c_{23}(\cos\delta - \cos(E_{12}L-\delta) - \cos(E_{13}L+\delta) + \cos(E_{23}L+\delta)) \,.	
	 \label{eq:nu_e_to_nu_mu}
\end{align}
When the torsional interaction is switched off, this agrees with the formula given in~{\cite{ParticleDataGroup:2022pth,Akhmedov:2004ny, Nunokawa:2005nx, Minakata:2006gq, Zaglauer:1988gz, Freund:2001pn, Kimura:2002wd, Cervera:2000kp} }. Similarly, the $\nu_e \to \nu_\tau$ 
conversion probability is
\begin{align}
P({\nu_e \to \nu_\tau})=&\frac{2\alpha^2}{2\tilde{A}^2}{s}_{12}^2 {c}_{12}^2 {s}_{23}^2 \left( 1-\cos({E_{12}L}) \right) +2 \frac{{s}_{13}^2}{(\tilde{A}-1)^2} {c}_{23}^{2}\left( 1-\cos({E_{13}L}) \right) \nonumber \\
	&+\frac{2\alpha {s}_{13}}{\tilde{A}(\tilde{A}-1)}s_{12}c_{12}s_{23}c_{23}\left(\cos(E_{12}L-\delta) +  \cos(E_{13}L+\delta) - \cos(E_{23}L+\delta)-\cos\delta \right) \,,
	 \label{eq:nu_e_to_nu_tau}
\end{align}
while the $\nu_e$ survival probability is
\begin{align}
P({\nu_e \to \nu_e})=&1-\frac{2\alpha^2}{2\tilde{A}^2}{s}_{12}^2 {c}_{12}^2  \left( 1-\cos({E_{12}L}) \right) - 2 \frac{{s}_{13}^2}{(\tilde{A}-1)^2} \left( 1-\cos({E_{13}L}) \right) \,.
	 \label{eq:nu_e_to_nu_e}
\end{align}}
For antineutrinos, the following replacements are necessary in {Eq. (\ref{eq:TDSE_for_3nu.b}),
\begin{align}
	U^* \to U ,\quad D \to -D, \quad \text{and} \quad \Delta \tilde{m}^2_{ij} := \Delta m^2_{ij} - 2 \tilde{n} E \Delta\lambda_{ij}\,.
\end{align}
We have calculated the anti-neutrino conversion and survival probabilities using these replacements in Eq.~(\ref{eq:TDSE_for_3nu_2}).
These match with the formulae given in~\cite{ParticleDataGroup:2022pth,Akhmedov:2004ny, Nunokawa:2005nx, Kimura:2002wd, KamLAND:2013rgu, DayaBay:2018yms} when the geometrical interaction is absent, and thus the effect of this interaction can be directly read off.

For example, the $\bar{\nu}_e \to \bar{\nu}_\mu$ conversion probability is given by 
{\begin{align}
P({\bar{\nu_e} \to \bar{\nu_\mu}})=&\frac{\alpha^2}{2\tilde{A}^2} \sin^2 2\theta_{12}{c}_{23}^2 \left( 1-\cos({E_{12}L}) \right) +2 \frac{{s}_{13}^2}{(\tilde{A}+1)^2} {s}_{23}^{2}\left( 1-\cos({E_{13}L}) \right) \nonumber \\
	&+\frac{2\alpha }{\tilde{A}(\tilde{A}+1)}{s}_{13}s_{12}c_{12}s_{23}c_{23}(\cos\delta - \cos(E_{12}L+\delta) - \cos(E_{13}L-\delta) + \cos(E_{23}L-\delta)) \,.	
	 \label{eq:nubar_e_to_nubar_mu}
\end{align}}
\begin{figure}[thbp]
	\includegraphics[width=5cm]{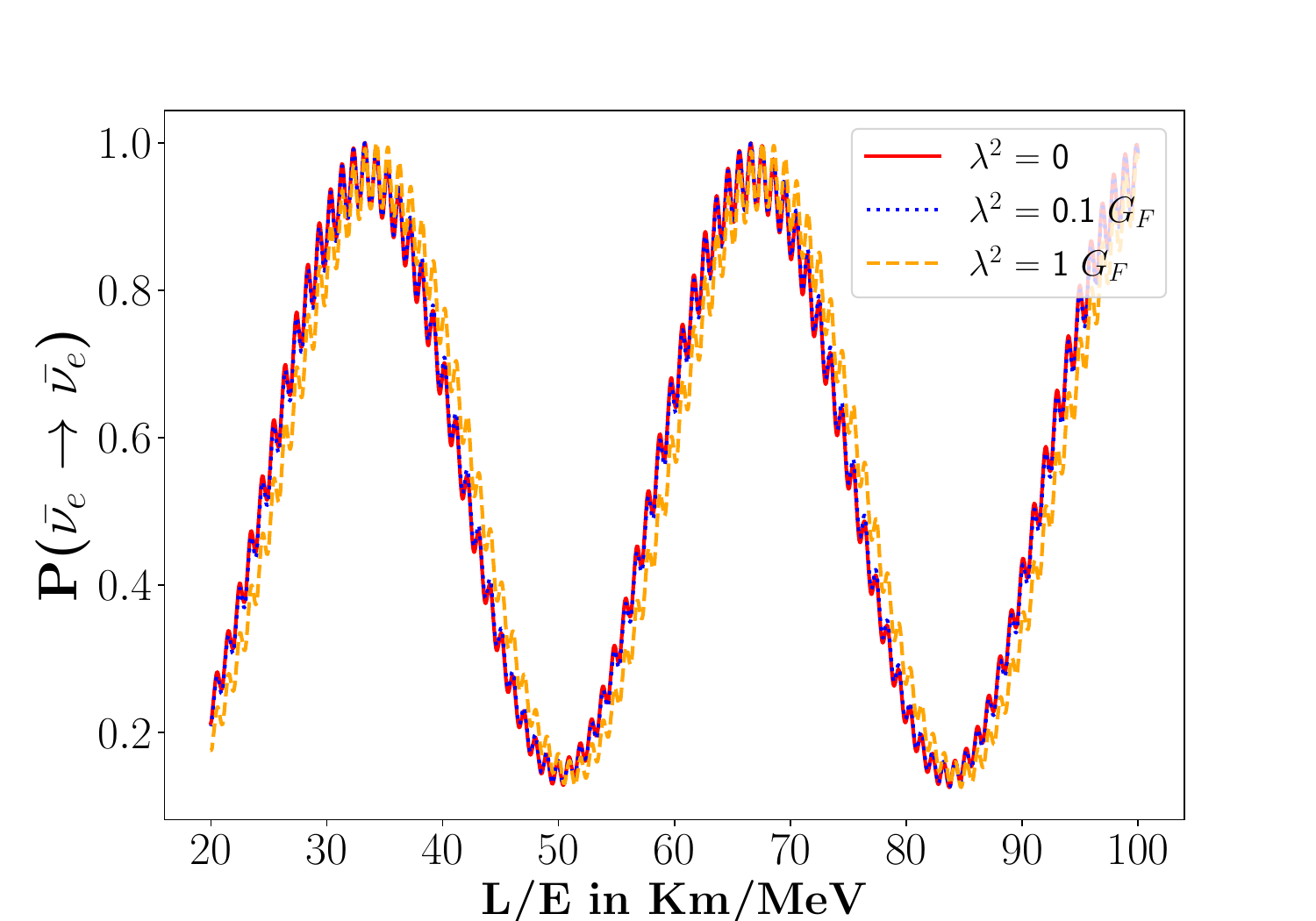}
	\caption{{$\bar{\nu}_e$ survival probability as a function of $L/E$\,, for a baseline $L=180\,km\,.$ 
			} 
	}
	\centering
	\label{anti-nu-e-3}
\end{figure}
We have also numerically calculated the $\bar{\nu_e}$ disappearance probability for different baselines. In Fig.~\ref{anti-nu-e-3}, 
we have plotted the survival probability of $\bar\nu_e$ for a baseline of 180 km, which corresponds to the KamLAND (LBL reactor neutrino) 
experiment. The plots for different scales of the geometrical coupling, $\lambda^2=0.1 G_F$ and $\lambda^2 = 1 G_F$\,, are shown in the
same figure as the case $\lambda=0\,.$ The effect of the geometrical interaction appears much smaller than in the higher baseline experiment, 
which is not surprising, as the effect decreases with decreasing baseline. However, an interesting consequence of the interaction appears when 
we plot the $\bar{\nu}_e$ survival probability against $L/E$ for two different baselines on the same plot, as we have done for $L= 53$ km (JUNO) 
and $L=180$ km (KamLAND) in Fig.~\ref{0.1,1gf-juno-kamland}. 
The probability now depends on $L$ and $E$ separately, which means that the curves for different baselines do not match in the overlap region
when plotted against $L/E\,.$
\begin{figure}[htbp]    
	\includegraphics[width=5cm]{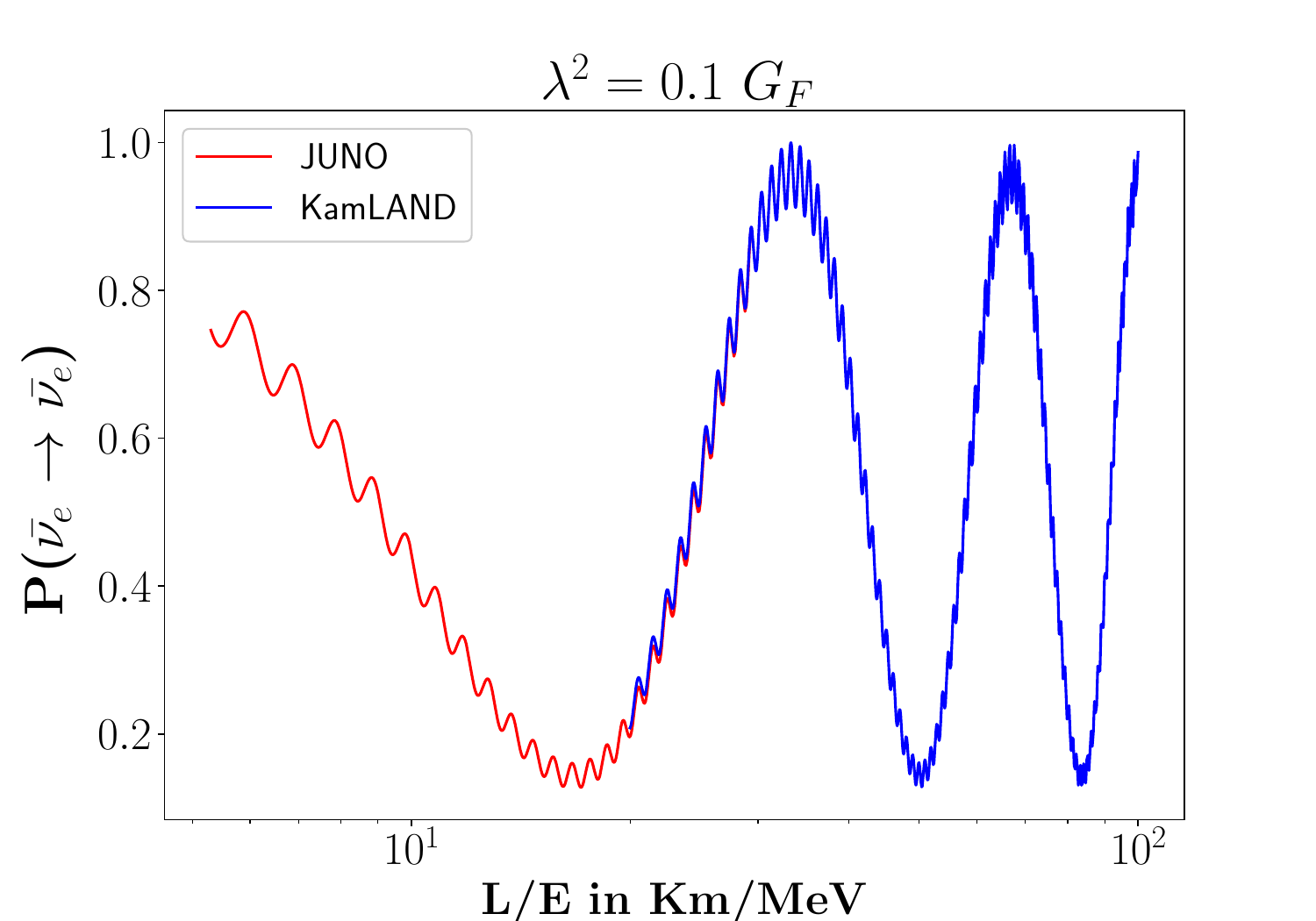}
	\includegraphics[width=5cm]{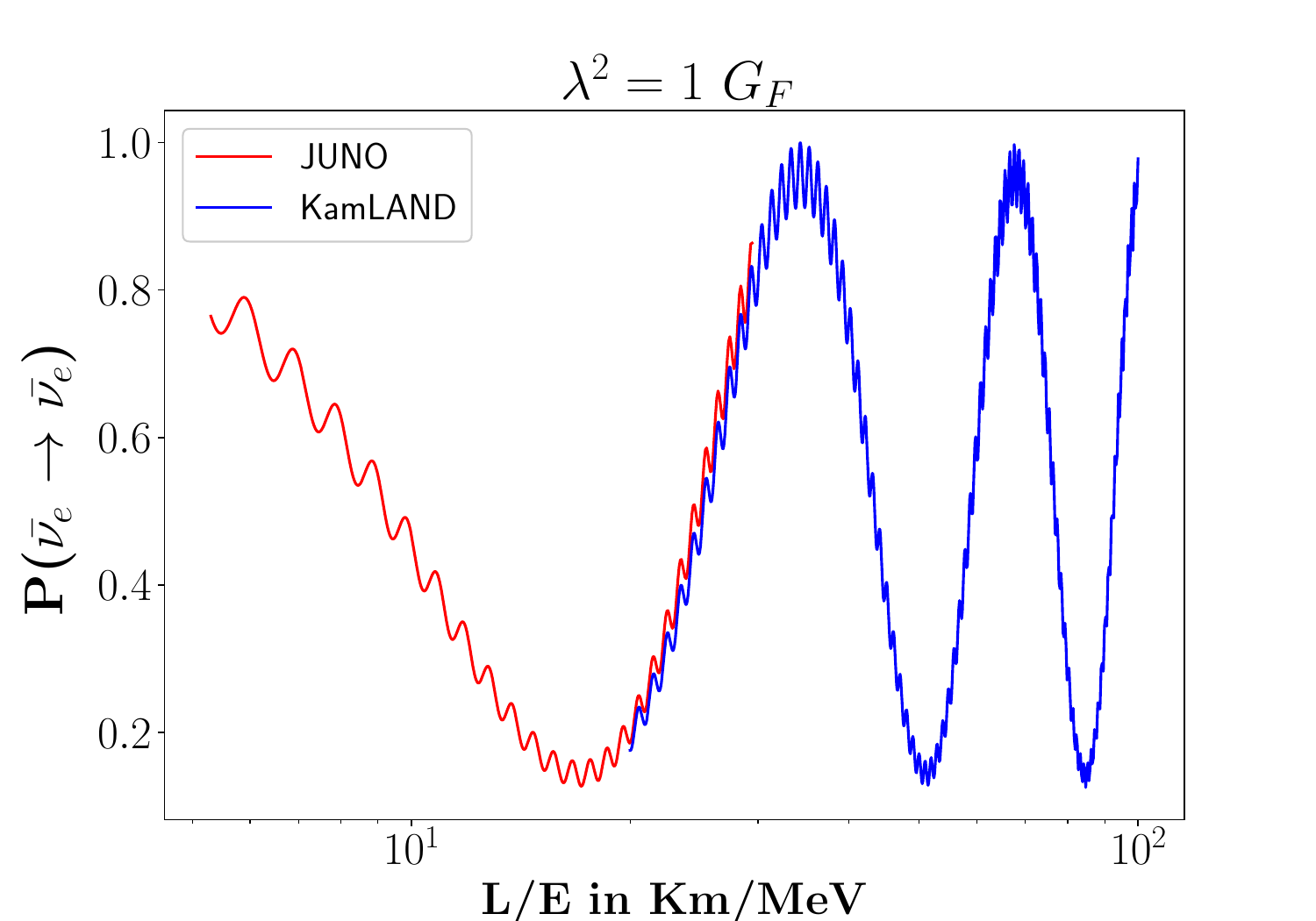}
		\caption{{$\bar{\nu}_e$ survival probability as a function of $L/E$ for $\lambda^2 = 0.1 G_F$ and $\lambda^2 = 1 G_F$ for  JUNO and KamLAND.}}
		\label{0.1,1gf-juno-kamland}
\end{figure}

\section{Discussion}\label{disc}
Our results clearly show that neutrino oscillations are affected by the geometrical coupling constants. Let us now try to understand how much of an effect we should expect, as we have avoided discussing the size of the coupling constants $\lambda$\, so far. 
Since contorsion appears as one part of the spin connection, with the other part contributing to 
gravitation, it is quite common to think of contorsion as part of gravity, {thus it is a often believed that the coupling constants should be suppressed by factors of $\frac{1}{M_{Pl}}$\,. This line of reasoning is not correct -- it can be seen quite easily that it is not possible to determine the size of the $\lambda$ from purely theoretical considerations, as we discuss now. }

{
Firstly, we note that the contorsion field, being non-dynamical, does not have a natural scale. So it can be eliminated at every scale, classically, without introducing a corresponding scale in the theory. This is in contrast to most field theories, where removing a field from the theory produces a coupling constant related to the mass of that field. For example,  effective four-fermion interactions typically arise from massive gauge boson exchange when the gauge boson propagator  $\frac{-g_{\mu\nu}}{q^2 - M^2}$ is replaced by $\frac{g_{\mu\nu}}{M^2}$ at low energies. This is what happens when the weak interactions are approximated by Fermi's theory of neutron decay so that the four-fermion coupling goes like $\frac{1}{M_W^2}$\,. However, the geometrical four-fermion interaction discussed in this paper is not a low-energy approximation and does not appear from a propagator -- the contorsion field does not have a propagator, it is not even a quantum field. Thus there is no  mass scale specific to the coupling constants $\lambda^{V,A}_f$\,. The $M_P$ that was absorbed in the $\lambda$ in Eq.~(\ref{V-A}) is not the ``scale'' of the geometrical interactions, but only sets the dimensionality of $\lambda$ as being of mass dimension $-1$\,. It does not imply that $\lambda$ is small compared to $1/M_W$ -- the dimensionless $\lambda$ of Eq.~(\ref{4fermi}) can be arbitrarily large, so multiplying them by $\sqrt{\kappa}$ need not make them negligibly small. }

Secondly, we are not doing quantum gravity --- quantization of spacetime is not required for what we have done --- {so $M_P$\,,} the scale of quantum gravity, is not ``natural'' in this interaction. The contorsion field is independent of the gravitational part of the spin connection,
as can be seen by writing the spin connection as $A_\mu{}^{ab} = \Lambda_\mu{}^{ab} + \omega_\mu{}^{ab}$\,, where the ``Levi-Civita spin connection'' $\omega_\mu{}^{ab}$ is built purely out of tetrads, co-tetrads and their derivatives, and corresponds to the unique torsion-free Levi-Civita connection in the metric formalism. So, even if there is a quantum theory of gravity and if the scale of that is indeed the Planck scale, it is not necessary that the scale of the contorsion field will be the same. Another way of saying this is the following. While the $\lambda_f$ are of mass dimension $-1$\,, we cannot find their values theoretically in the absence of a quantum theory of spacetime which produces Eq.~(\ref{L_psi_all}) at low energies. There is also no symmetry which forces the $\lambda_f$ to vanish, so they should be included in calculations and experimental results used to set bounds on them. {An analogy would be the various Yukawa couplings, which cannot be fixed purely by theoretical arguments, but require the observed fermion masses. The analogy is imperfect because the Yukawa couplings, being dimensionless, cannot be larger than unity for electroweak perturbation theory to hold, but there is no such restriction on the $\lambda\,$ of Eq.~(\ref{4fermi}).}

The geometrical four-fermion interactions are very similar to the non-standard neutrino interactions (NSI) known in the literature~\cite{Ohlsson:2012kf,Blennow:2016etl,Biggio:2009nt,Denton:2022pxt}, but there are some differences. The NSI are interactions of neutrinos in the flavor space, while for the geometrical interactions, the neutrinos are in the mass basis. {It is of course possible to rewrite the Hamiltonian in either basis --- a diagonal Hamiltonian in one basis is non-diagonal in the other. However, we are not discussing general NSI in this paper, but a specific kind of four-fermion interaction which applies to all fermions and not only neutrinos, and which is inevitably generated by the fact that spacetime is not flat. A general non-diagonal NSI will have many more parameters and is not the same as what we are discussing. Similarities with specific NSI models such as~\cite{Ge:2018uhz, Medhi:2021wxj} are accidental --- any four-fermion interaction involving all fermions would be similar in some sense.}

Another important distinction is that NSI usually assumes fields and interactions beyond the Standard Model, even if they are not made explicit in the low energy effective theory. That is not required for the geometrical interactions, as the four-fermion terms appear naturally from considering the dynamics of fermions in curved spacetime, {which guarantees} that all fermions are involved in these four-fermion interactions.  
{In the absence of flavor mixing, $e_L$ and $\nu_{eL}$ should have the same geometrical coupling $\lambda_{eL}$. That is not true when neutrino mixing occurs. However, for the purpose of estimating the effect of the geometrical interaction, we have assumed that $\lambda$ is of the same order of magnitude for electrons and neutrinos, and the differences in the values of $\lambda$ for different neutrinos are also of the same order of magnitude.  

Finally, since the mechanism behind neutrino masses is undetermined, we have also looked at the conversion probabilities for $\Delta m^2_{ij}=0\,,$ but with $\Delta\lambda^2 = 0.1G_F\,$ and $1G_F\,.$  For the baseline of $1300~km\,,$  these probabilities are vanishingly small, though not actually zero. They will be larger for very long baselines and thus may become relevant in future measurements in such experiments~\cite{Gandhi:2004bj,Gandhi:2007td}. We will discuss this elsewhere. 

\begin{acknowledgments}
	The authors thank P.~B.~Pal, A.~Raychaudhuri, and T.~Schwetz for helpful comments.
\end{acknowledgments}

\bigskip

{\bf Data availability statement:} This manuscript has no associated data
or the data will not be deposited. [Authors’ comment: There are no data
because this is a purely theoretical work.]

\end{document}